\documentclass[12pt,preprint]{aastex}

\def\kms{\,km\,s$^{-1}$} %kms -1 
\def\vsini{$v\sin i$}
\begin{document}

\title{Testing Stellar Models With An
	Improved Physical Orbit for 12~Bo\"{o}tis}
\author{Andrew F.~Boden\altaffilmark{1,2},
	Guillermo~Torres\altaffilmark{3},
	Christian A.~Hummel\altaffilmark{4}
}
\altaffiltext{1}{Michelson Science Center, California Institute of Technology}
\altaffiltext{2}{Department of Physics and Astronomy, Georgia State University}
\altaffiltext{3}{Harvard-Smithsonian Center for Astrophysics}
\altaffiltext{4}{European Southern Observatory}
\email{bode@ipac.caltech.edu}

\begin{abstract}
In a previous publication we reported on the binary system
12~Bo\"{o}tis and its evolutionary state.  In particular the 12~Boo
primary component is in a rapid phase of evolution, hence accurate
measurement of its physical parameters makes it an interesting test
case for stellar evolution models.  Here we report on a significantly
improved determination of the physical orbit of the double-lined
spectroscopic binary system 12~Boo.  We have a 12~Boo interferometry
dataset spanning six years with the Palomar Testbed Interferometer, a
smaller amount of data from the Navy Prototype Optical Interferometer,
and a radial velocity dataset spanning 14 years from the
Harvard-Smithsonian Center for Astrophysics.  We have updated the
12~Boo physical orbit model with our expanded interferometric and
radial velocity datasets.  The revised orbit is in good agreement with
previous results, and the physical parameters implied by a combined
fit to our visibility and radial velocity data result in precise
component masses and luminosities.  In particular, the orbital
parallax of the system is determined to be 27.74 $\pm$ 0.15 mas, and
masses of the two components are determined to be 1.4160 $\pm$ 0.0049
M$_{\sun}$ and 1.3740 $\pm$ 0.0045 M$_{\sun}$, respectively.  These
mass determinations are more precise than the previous report by a
factor of four to five.

As indicated in the previous publication, even though the two
components are nearly equal in mass, the system exhibits a significant
brightness difference between the components in the near infrared and
visible.  We attribute this brightness difference to evolutionary
differences between the two components in their transition between
main sequence and giant evolutionary phases, and based on theoretical
models we can estimate a system age of approximately 3.2 Gyr.
Comparisons with stellar models suggest that the 12~Boo primary may be
just entering the Hertzsprung gap, but that conclusion is highly
dependent on details of the models.  Such a dynamic evolutionary state
makes the 12~Boo system a unique and important test for stellar
models.

\end{abstract}


\section{Introduction}

12~Bo\"{o}tis (d~Bo\"{o}tes, HR~5304, HD~123999, HIP~69226) is a
short-period (9.6 d) binary system with nearly-equal mass ($q \sim$
0.97) components.  The system was first detected as a radial velocity
variable over 100 years ago \citep{Campbell1900}, and the first
``good'' double-lined orbit was calculated by \citet{Abt76}.  The
\citet{Abt76} orbit has been reconfirmed by an independent CORAVEL
radial velocity orbit by \citet[hereafter DU99]{DeMedeiros99}
\citep[data from which was used in][]{Boden2000}.  The composite
system has been consistently assigned the spectral type F8IV -- F9IVw,
the latter by \citet{Barry70}, with the ``w'' indicating weak
ultraviolet metallic features.  All studies seem to confirm that
\objectname[HD 123999]{12~Boo} has heavy element abundances near solar
proportions
\citep{Duncan81,Balachandran90,Lebre99,Nordstrom2004}.

Previously we reported a physical orbit model for the 12~Boo system
\citep[hereafter Paper~1]{Boden2000}.  Paper~1 discussed the
interesting evolutionary state of the 12~Boo components; despite the
nearly equal mass ratio, the 12~Boo components exhibit an unusual
intensity ratio due to their positions on the Hertzsprung-Russel
diagram.  However, the orbit model of Paper~1 relied upon rather
limited radial velocity data.  Given this shortcoming and the
favorable geometry of the 12~Boo system for high-precision study, we
decided to refine the orbit model in order to fully exploit the 12~Boo
components as a test of stellar models.  Consequently, herein we
report on a significantly improved determination of the 12~Boo
physical orbit from an expanded set of near-infrared, long-baseline
interferometric measurements taken with the Palomar Testbed
Interferometer (PTI) and Navy Prototype Optical Interferometer (NPOI),
and a large set of new spectroscopic radial velocity measurements
obtained at the Harvard-Smithsonian Center for Astrophysics (CfA).

In the following we discuss the new observations
(\S\ref{sec:observations}), the orbit model (\S\ref{sec:orbit}) and
physical properties (\S\ref{sec:physics}) derived from them, and
compare the component properties with stellar models
(\S\ref{sec:models}).  We summarize our findings in
\S\ref{sec:summary}.

\section{Observations}
\label{sec:observations}

\paragraph{Interferometry}
As in Paper~1, the interferometric observable used for these
measurements is the fringe contrast or {\em visibility} (squared) of
an observed brightness distribution on the sky.  PTI was used to make
the interferometric measurements presented here; PTI is a
long-baseline $H$ (1.6$\mu$m) and $K$-band (2.2$\mu$m) interferometer
located at Palomar Observatory, and described in detail elsewhere
\citep{Colavita99a}.  The analysis of such data in the context of a
binary system is discussed in detail in Paper~1 and elsewhere
\citep[e.g.][]{Hummel2001} and will not be repeated here.

12~Boo was observed in conjunction with objects in our calibrator list
by PTI in $K$ band ($\lambda \sim 2.2 \mu$m) on 67 nights between 21
June 1998 and 18 June 2004, a dataset covering roughly six years and
228 orbital periods.  Additionally, 12~Boo was observed by PTI in
$H$ band ($\lambda \sim 1.6 \mu$m) on 12 nights between 28 May 1999
and 15 June 2001.  12~Boo, along with calibration objects, was usually
observed multiple times during each of these nights, and each
observation, or scan, was approximately 130 sec long.  For each scan
we computed a mean $V^2$ value from the scan data, and the error in
the $V^2$ estimate from the rms internal scatter \citep{Colavita99b}.
12~Boo was always observed in combination with one or more calibration
sources within $\sim$ 10$^{\circ}$ on the sky.  As in Paper~1, here we
have used three stars as calibration objects: \objectname[HD
121107]{HD~121107} (G5 III), \objectname[HD 128167]{HD~128167} (F2 V),
and \objectname[HD 123612]{HD~123612} (K5 III).  Table
\ref{tab:calibrators} lists the relevant physical parameters for the
calibration objects.

\begin{deluxetable}{ccccc}
\tabletypesize{\small}
\tablecolumns{5}
\tablewidth{0pc}

\tablecaption{PTI 12~Boo $V^2$ Calibration Objects Considered in our
Analysis.  The relevant parameters for our three calibration objects
are summarized.  Apparent diameter values are determined from spectral
energy distribution modeling of archival broad-band photometry and
spectral typing, and visibility measurements with PTI.
\label{tab:calibrators}
}

\tablehead{
\colhead{Object} & \colhead{Spectral} & \colhead{Star}      & \colhead{Separation} & \colhead{Adopted Model} \\
\colhead{Name}   & \colhead{Type}     & \colhead{Magnitude} & \colhead{from 12~Boo}& \colhead{Diameter (mas)}
}

\startdata
HD 121107 & G5 III   & 5.7 $V$/4.0 $K$ & 8.2$^{\circ}$  & 0.83 $\pm$ 0.06   \\
HD 128167 & F2 V     & 4.5 $V$/3.5 $K$ & 7.1$^{\circ}$  & 0.77 $\pm$ 0.04   \\
HD 123612 & K5 III   & 6.6 $V$/3.1 $K$ & 0.92$^{\circ}$ & 1.29 $\pm$ 0.10   \\
\enddata

\end{deluxetable}

The calibration of 12~Boo $V^2$ data is performed by estimating the
interferometer system visibility ($V^{2}_{sys}$) using calibration
sources with model angular diameters, and then normalizing the raw
12~Boo visibility by $V^{2}_{sys}$ to estimate the $V^2$ measured by
an ideal interferometer at that epoch \citep{Mozurkewich91,Boden98}.
Calibrating our 12~Boo dataset with respect to the three calibration
objects listed in Table \ref{tab:calibrators} results in a total of
303 calibrated scans (258 in $K$, 46 in $H$) on 12~Boo over 78 nights,
roughly quadrupling the visibility dataset from Paper~1.  Our
calibrated synthetic wide-band $V^2$ measurements are summarized in
Table~\ref{tab:V2Table}.  In particular, all $V^2$ data from
Paper~1 are also contained in Table~\ref{tab:V2Table} and used in this
analysis.  Table~\ref{tab:V2Table} gives $V^2$ measurements and times,
measurement errors, residuals between our data and orbit model
(Table~\ref{tab:orbit}) predictions, the photon-weighted average
wavelength, $u-v$ coordinates, and on-target hour angle for each of
our calibrated PTI 12~Boo $V^2$ observations.

\begin{table}
\dummytable\label{tab:V2Table}
\end{table}

In addition to the PTI visibility data, we have obtained new
visibility and closure-phase data on 12~Boo with the Navy Prototype
Optical Interferometer \citep[NPOI;][]{Armstrong1998,Hummel2003}.
NPOI observed 12~Boo and calibrator HD~128167
(Table~\ref{tab:calibrators}) on 13 nights from 9 April 2001 through
28 June 2001 inclusive.  On five nights, the NPOI C, E, and W stations
were used with maximum baseline of 38~m, and on all other nights the
W7, E, and W stations with a maximum baseline of 64~m.  The data were
taken in 10 narrow-band channels spanning a passband between 650 nm
and 850 nm.  Unlike the single-baseline PTI $V^2$ data, NPOI data were
all taken simultaneously using three NPOI baselines; in addition to
visibility amplitude such data provide phase-closure information.  In
particular the NPOI phase-closure data breaks the $\Omega$/inversion
degeneracy inherent in the PTI $V^2$ (see \S~\ref{sec:orbit}).  The
NPOI data were reduced and calibrated according to the standard
procedures outlined by \citet{Hummel1998}.

\begin{table}
\dummytable\label{tab:RVdata}
\end{table}

\paragraph{Spectroscopy}

The largest shortcoming in the orbit model from Paper~1 was the
limited and inhomogeneous radial velocity (RV) data.  To improve this
situation we have obtained 49 new high-resolution spectra of the
12~Boo system at CfA, with an \'{e}chelle spectrograph mounted at the
Cassegrain focus of the 1.5-m Wyeth reflector at the Oak Ridge
Observatory (Harvard, Massachusetts).  The resolving power of this
instrument is $\lambda/\Delta\lambda \approx 35,000$. A single
\'{e}chelle order spanning 45~\AA\ was recorded with a photon-counting
Reticon detector, at a central wavelength of 5188.5~\AA, near the
\ion{Mg}{1}~b triplet.
%% The photon
%%counts per resolution element of 8.5~\kms\ range from about 100 to
%%3100. However, at the higher count levels the signal-to-noise ratio
%%actually achieved is limited by shortcomings in the flat-fielding
%%rather than in photon statistics and rarely exceeds 50.
12~Boo observations were carried out between June 1987 and April 2001,
spanning nearly 14 years and 526 system periods.

Component velocities were determined from the spectra using the TODCOR
two-dimensional cross-correlation technique \citep{ZM94}, with
synthetic templates for the each star based on Kurucz model
atmospheres (available at http://cfaku5.cfa.harvard.edu). The optimal
parameters for the templates (mainly effective temperature and
rotational velocity) were determined by seeking the best match to the
observed spectra as judged from the peak cross-correlation value
averaged over all exposures.  We obtained $v \sin i$ values of
14~\kms\ and 12~\kms\ for the primary (more massive star) and
secondary, respectively, with estimated errors of 1~\kms.  The
effective temperatures are sensitive to the adopted surface gravity
and chemical composition.  For estimated $\log g$ values of 3.8 and
4.0 (see \S~\ref{sec:physics}), we obtained by interpolation
temperatures of 6130~K and 6230~K for the primary and secondary with
adopted uncertainties of 100~K and 150~K, respectively.  These are in
good agreement with photometric estimates for 12~Boo (\S
\ref{sec:physics}).  Tests with adopted metallicities between [m/H]$=
-1.0$ and [m/H]$= +0.5$ in steps of 0.5 dex indicated a preference for
solar composition, consistent with recent estimates from the
literature.  We also determined the light ratio from our spectra
following \cite{ZM94}. The result is $L_{\rm sec}/L_{\rm prim} = 0.64
\pm 0.02$ at the mean wavelength of our observations, which is close
to the visual band.
%%A similar value was obtained from the NPOI visibility data.

The radial velocities were tested for systematics by performing
numerical simulations following \citet{Torres2002}, and small
adjustments (typically less than 0.5~\kms) were applied to the raw
velocities as corrections. The stability of the zero point of our
velocity system was monitored by means of exposures of the dusk and
dawn sky, and small systematic run-to-run corrections were applied in
the manner described by \citet{Latham1992}.  The final radial
velocities are given in Table~\ref{tab:RVdata}, which gives measured
primary and secondary velocities and associated uncertainties,
residuals between the measurements and predictions from our orbit
model (Table~\ref{tab:orbit}), and model phase.

\section{Orbit Model}
\label{sec:orbit}
As in previous papers in this series
\citep{Boden99a,Boden99b,Boden2000,BL2001,Torres2002}, the estimation
of the 12~Boo orbit is made by fitting a Keplerian orbit model
directly to the calibrated PTI $V^2$ and RV data on 12~Boo \citep[see
also][]{Armstrong92b,Hummel1993,Hummel1995,Hummel1998,Hummel2001}.  As
discussed in these references, the nature of $V^2$ data require that
orbital parameter estimation necessarily include in-band component
intensity ratios and component diameters (diameters here were held
fixed at photometrically estimated values; see \S\ref{sec:physics}).
Given the well-established 12~Boo model from Paper~1, the additional
data presented here are straightforwardly used to refine the previous
orbit model.

In general the NPOI data were found to agree well with the orbit
estimated from the PTI $V^2$ and RV data.  However with the exception
of the closure phases these data did little to further constrain the
orbit parameters.  Therefore we have adopted the general orbit
orientation (i.e.~$\Omega$) indicated by the NPOI closure phases, but
the high-precision orbital parameters are derived solely from the PTI
$V^2$ (Table~\ref{tab:V2Table}) and RV (Table~\ref{tab:RVdata}) data.

\begin{figure}
\epsscale{0.7}
%%\plotone{figures/12boo.trace.eps}
\plotone{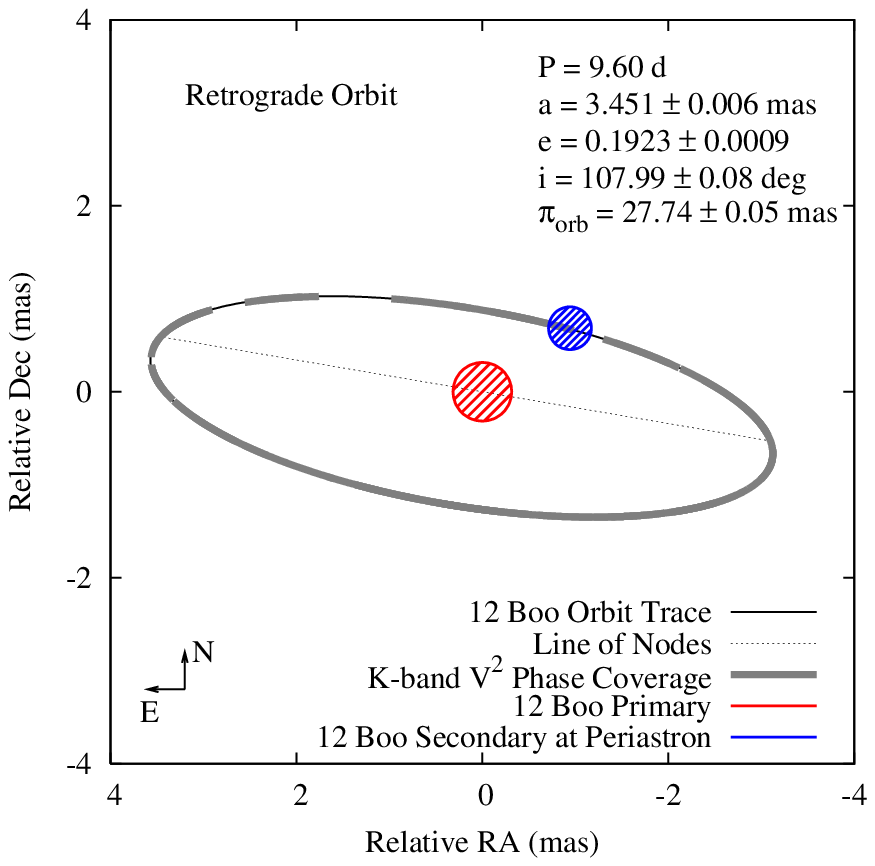}
\caption{Visual Orbit of 12~Bo\"{o}tis.  The relative visual orbit
model of 12~Boo is shown, with the primary and secondary objects
rendered at T$_0$ (periastron).  The heavy lines along the relative
orbit indicate areas where we have phase coverage in our $K$-band PTI
data (they are not separation vector estimates); our data cover
essentially all phases of the orbit, leading to a reliable orbit
determination.  Component diameter values are estimated (see
discussion in \S~\ref{sec:physics}), and are rendered to scale.
\label{fig:12boo_orbit}}
\end{figure}

Figure \ref{fig:12boo_orbit} depicts the relative visual orbit of the
12~Boo system, with the primary component rendered at the origin, and
the secondary component rendered at periastron.  We have indicated the
phase coverage of our $V^2$ data on the relative orbit with heavy
lines; our data cover essentially all phases of the orbit, leading to
a reliable orbit determination.  Note that relative to Paper~1 the
orbit is inverted around the origin; the $V^2$ data used in Paper~1
and here are invariant under a mirror reflection of the component
relative positions, thus they does not distinguish between the two
orbit orientations.  The $V^2$ observable degeneracy was noted in
Paper~1 (in particular see the notes to Paper~1 Table~5), and is
broken by the addition of the closure phase data from NPOI.  Figure
\ref{fig:12boo_orbit} thus depicts the 12~Boo orbit as it appears on
the sky.

\clearpage
%% Figure 1

%% Figure 2
\begin{figure}
\epsscale{1.0}
%%\plotone{figures/12boo.rvPhase.eps}
\plotone{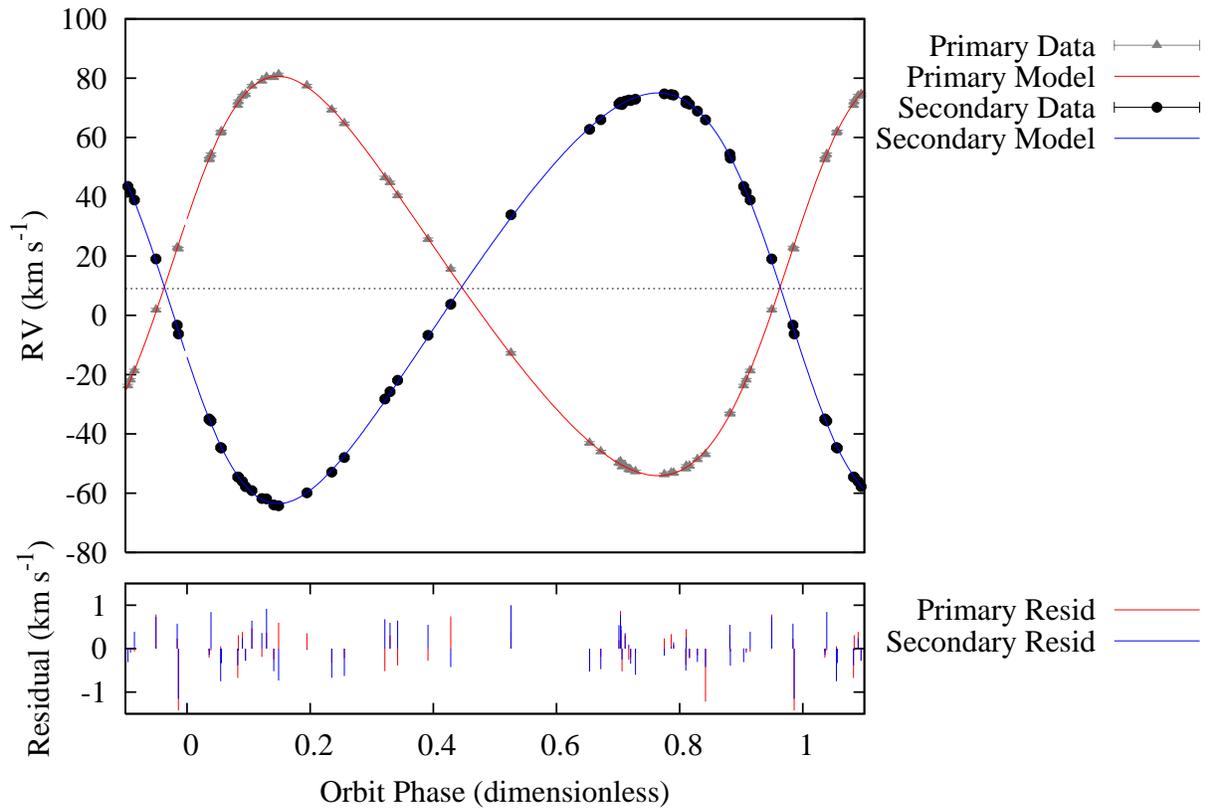}
\caption{Radial Velocity Orbit of 12~Boo.  A phase-folded plot of our
radial velocity data and RV predictions from our Full-Fit solution
(Table~\ref{tab:orbit}).  The lower frame gives RV residuals to the
model fit.
\label{fig:12boo_RVfit}}
\end{figure}
\clearpage

\begin{figure}
\epsscale{0.8}
%%\plotone{figures/12boo.v2Resid.eps}\\
%%\plotone{figures/12boo.rvResid.eps}
\plotone{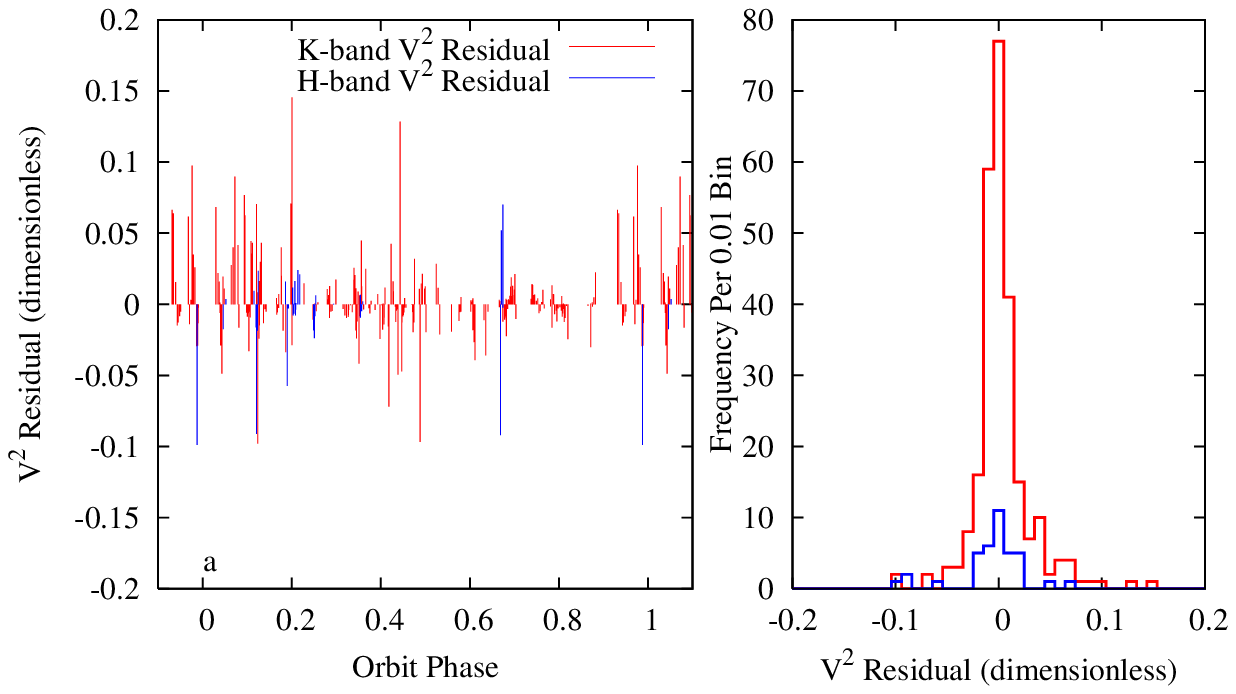}\\
\plotone{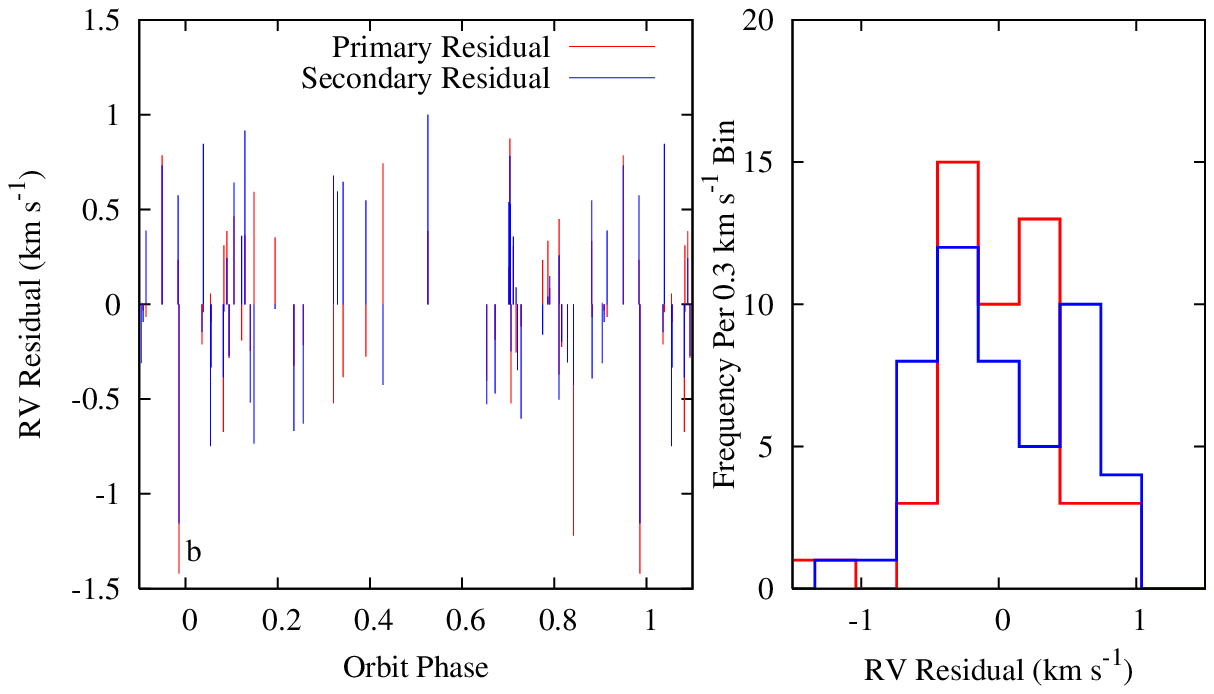}
\caption{Orbit Fit Residuals for 12~Boo.  a: Orbit phase plots of $K$
and $H$-band V$^2$ fit residuals, and residual histograms for the
``Full-Fit'' orbit model.  b: Orbit phase plots of radial velocity fit
residuals, and residual histograms for the ``Full-Fit'' orbit model.
(Table~\ref{tab:orbit}).
\label{fig:12boo_fitResiduals}}
\end{figure}

Tables \ref{tab:V2Table} and \ref{tab:RVdata} list the constituent set
of $V^2$ and RV measurements in our 12~Boo dataset, and residuals (in
a datum minus model sense) between the observables and predictions
based on the best-fit integrated orbit model (our ``Full-Fit'' model,
Table~\ref{tab:orbit}) for 12~Boo.  Figures \ref{fig:12boo_RVfit} and
\ref{fig:12boo_fitResiduals} illustrate the results of our orbit
modeling for 12~Boo.  Figure \ref{fig:12boo_RVfit} depicts phased RV
measurements and the primary and secondary radial velocity orbits from
our integrated model.  Inset in the lower frame are phased velocity
residuals (data minus model).  Figure \ref{fig:12boo_fitResiduals}
depicts the phase coverage of our visibility and radial velocity data,
and the statistics of our modeling residuals.  The agreement between
our various data and our orbit model is excellent; our full-fit
solution results in total chi-squared per of degree of freedom in our
fit of 0.85 (suggestive that we may have overestimated our measurement
errors).  The resulting rms $V^2$ and RV measurement residuals from
our model are 0.027 and 0.49 km s$^{-1}$ respectively.

\clearpage
%% Figure 3
\clearpage

Orbit models for 12~Boo are summarized in Table~\ref{tab:orbit},
including spectroscopic orbit parameters from DU99, the integrated
model from Paper~1, and our present visual, spectroscopic, and
integrated orbit models.  In particular we list the results of
separate fits to only our $K$-band $V^2$ data (our ``$V^2$ Only''
solution), our double-lined radial velocity data (our ``RV Only''
solution), and a simultaneous fit to our $V^2$ and RV data (our
``Full-Fit'' solution) -- all with component diameters constrained as
noted above.  For the orbit parameters we have estimated from our
visibility data we list a total one-sigma error in the parameter
estimate, including errors in the parameter estimates from statistical
(measurement uncertainty) and systematic error sources.  In our
analysis the dominant forms of systematic error are: (1) uncertainties
in the calibrator angular diameters (Table \ref{tab:calibrators}); (2)
uncertainty in the center-band operating wavelength ($\lambda_0
\approx$ 2.2 $\mu$m), taken to be 10 nm ($\sim$0.5\%); (3) the
geometrical uncertainty in our interferometric baseline ( $<$ 0.01\%);
and (4) uncertainties in ancillary parameters constrained in our orbit
fitting procedure (i.e.~the angular diameters in all solutions
involving interferometry data).  For example, our angular semi-major
axis error is completely dominated by the operating wavelength
uncertainty -- the statistical error is a factor of five smaller.

In addition to the overall orientation, the NPOI data have been used
to estimate the component in-band intensity ratio, constraining to the
relevant orbital parameters from the ``Full Fit'' solution.  The 750
nm intensity ratio is constrained by the NPOI closure phases and the
ratio of the maximum to minimum visibility amplitudes.  As there were
small systematic, but wavelength-independent deviations of the
measured amplitudes on individual baselines from the model, we allowed
the amplitude calibration to vary during the fitting (while the phases
are unaffected).  The resulting component intensity ratio (0.614 $\pm$
0.038) yields a magnitude difference $\Delta M$ = 0.53 $\pm$ 0.07 at
750 nm.  This value agrees well with the relatively increasing
secondary contribution seen in the other observed bands ($K$, $H$, and
$V$; component magnitude differences in all observed bands are given
in Table~\ref{tab:orbit}), and is understood in the context of the
slightly higher temperature of the 12~Boo secondary compared to the
primary (\S\ref{sec:observations}).

\begin{deluxetable}{l|cc|ccc}
\tabletypesize{\footnotesize}
\tablecolumns{6}
\tablewidth{0pc}
\rotate

\tablecaption{Orbital Parameters for 12~Boo.  Summarized here are the
apparent orbital parameters for the 12~Boo system as determined by
DU99, Paper~1, and present results.  We give three separate fits to
our data: $K$-band V$^2$ only, RV only, and integrated (``Full Fit'').
Note that $\omega_{A}$ refers to the argument of periapsis of the
primary component, and (unlike Paper~1) $\Omega$ values are the
position angle of the ascending node.  \label{tab:orbit}}

\tablehead{
\colhead{Orbital}   & \colhead{DU99} & \colhead{Paper~1} & \multicolumn{3}{c}{PTI \& CfA} \\
\colhead{Parameter} &                & \colhead{Full-Fit} & \colhead{$K$-band V$^2$ Only} & \colhead{RV Only} & \colhead{Full-Fit} }
\startdata
Period (d)          & 9.6046     & 9.604565              & 9.604638             & 9.6045518          & 9.6045492      \\
                    & $\pm$ 1 $\times$ 10$^{-4}$  & $\pm$ 1.0 $\times$ 10$^{-5}$ &$\pm$ 5.9 $\times$ 10$^{-5}$ & $\pm$ 8.6 $\times$ 10$^{-6}$ & $\pm$ 7.6 $\times$ 10$^{-6}$ \\
T$_{0}$ (MJD)       & 48990.29       & 51237.779         & 51237.7596           & 51237.7798         & 51237.7729 \\
                    & $\pm$ 0.03     & $\pm$ 0.024       &  $\pm$ 0.0086        & $\pm$ 0.0090       & $\pm$ 0.0051 \\
$e$                 & 0.193          &  0.1884           & 0.1895               & 0.19256            & 0.19233 \\
 		    & $\pm$ 0.004    & $\pm$ 0.0022      & $\pm$ 0.0022         & $\pm$ 0.00099      & $\pm$ 0.00086 \\
K$_1$ (km s$^{-1}$)     & 67.11 $\pm$ 0.41 & 67.84 $\pm$ 0.31 &                 & 67.320 $\pm$ 0.090 & 67.302 $\pm$ 0.087 \\
K$_2$ (km s$^{-1}$)     & 70.02 $\pm$ 0.48 & 69.12 $\pm$ 0.48 &                 & 69.38 $\pm$ 0.10   & 69.36 $\pm$ 0.10  \\
$\gamma$ (km s$^{-1}$)  & 9.29 $\pm$ 0.19  & 9.11 $\pm$ 0.13  &                 & 9.550 $\pm$ 0.051  & 9.551 $\pm$ 0.051 \\
$\omega_{A}$ (deg)      & 286.19 $\pm$ 1.31 & 287.03 $\pm$ 0.75 & 286.85 $\pm$ 0.35 & 286.92 $\pm$ 0.35 & 286.67 $\pm$ 0.19 \\
$\Omega$ (deg)          &            & 79.83 $\pm$ 0.45  & 80.49 $\pm$ 0.12   &                 & 80.291  $\pm$ 0.079 \\
$i$ (deg)               &            & 108.58 $\pm$ 0.36 & 108.15 $\pm$ 0.12    &                 & 107.990 $\pm$ 0.077 \\
$a$ (mas)               &            & 3.392 $\pm$ 0.050 & 3.449 $\pm$ 0.018    &                 & 3.451 $\pm$ 0.018 \\
$\Delta K_{\rm CIT}$ (mag)        &         & 0.618 $\pm$ 0.022 & 0.593 $\pm$ 0.006 &             & 0.589 $\pm$ 0.005 \\
$\Delta H_{\rm CIT}$ (mag)        &         & 0.566 $\pm$ 0.066 &&                                & 0.560 $\pm$ 0.020 \\
$\Delta M_{750 nm}$ (mag)         &         &                   &&                                & 0.53 $\pm$ 0.07  \\
$\Delta V$ (mag)        &         & 0.5 $\pm$ 0.1     &&                                          & 0.485 $\pm$ 0.017 \\
$\chi^2$/DOF	        &                   & 1.2   & 0.82 & 1.0 & 0.85 \\
$\overline{|R_{V^2}|}$/$\sigma_{V^2}$ &                   & 0.023 & 0.017/0.027 & & 0.017/0.027 \\
$\overline{|R_{RV}|}$/$\sigma_{RV}$ (km s$^{-1}$)  & 0.90 & 1.7  & & 0.39/0.49  & 0.40/0.49\\
\hline
\enddata

\end{deluxetable}

\section{Physical Parameters}
\label{sec:physics}
Physical parameters derived from our 12~Boo ``Full-Fit'' integrated
visual/spectroscopic orbit are summarized in Table \ref{tab:physics}.
Notable among these is the high-precision determination of the
component masses for the system, a virtue of the favorable geometry of
the orbit and the quality of the visibility and radial velocity
datasets.  We estimate the masses of the primary and secondary
components as 1.4160 $\pm$ 0.0049 and 1.3740 $\pm$ 0.0045 M$_{\sun}$,
respectively.  These are in good agreement (approximately 1.0 and 1.7
sigma for the primary and secondary respectively) with the component
mass estimates given in Paper~1.

The Hipparcos catalog lists the parallax of 12~Boo as 27.27 $\pm$ 0.78
mas \citep{HIP97}.  The distance determination to 12~Boo based on our
orbital solution is 36.08 $\pm$ 0.19 pc, corresponding to an orbital
parallax of 27.72 $\pm$ 0.15 mas, consistent with the Hipparcos
result at 1.7\% and 0.6-sigma.

A number of metallicity estimates for 12~Boo exist in the literature
that appear to indicate a composition near solar. Photometric
estimates by \cite{Duncan81}, \cite{Balachandran90}, and
\cite{Nordstrom2004} give [Fe/H] values of $-0.03$, $+0.12 \pm 0.12$,
and $-0.06$, respectively, and are based on Str\"omgren $uvby\beta$ or
$\delta(U-B)_{0.6}$ indices. Although the object was recognized as a
binary in these investigations, no corrections for this were made.
The effect is expected to be small in any case, as the two components
have very similar temperatures.  Spectroscopic determinations of the
metallicity have been reported by \cite{Balachandran90} and
\cite{Lebre99} as [Fe/H] = $-0.03 \pm 0.09$ and [Fe/H] = $-0.1 \pm
0.1$, respectively. Once again the binary nature of 12~Boo was known
to these investigators, although \cite{Balachandran90} reported not
detecting the secondary in their spectra. The effect would be to make
the spectral lines appear somewhat weaker, since the secondary
($L_{\rm sec}/L_{\rm prim} = 0.64$) would tend to fill in the lines of
the primary.  Overall there is good agreement in that all these
studies place the metallicity of 12~Boo within 0.1 dex of solar.  This
metallacity range is an important constraint we use below in the
comparison with stellar evolution models.

\begin{deluxetable}{ccc}
\tabletypesize{\small}
\tablecolumns{3}
\tablewidth{0pc}

\tablecaption{Physical Parameters for 12~Boo.  Summarized here are the
physical parameters for the 12~Boo system as derived primarily from
the ``Full-Fit'' solution orbital parameters in Table \ref{tab:orbit}.
For comparison we have also given the corresponding parameter values
from Paper~1 (offset in brackets and italics).  We have used system
photometry in $K$ and $H$ from Paper~1, and in $V$ from
\citet{Mermilliod1994}.
\label{tab:physics}}

\tablehead{
\colhead{Physical}   & \colhead{Primary (A)}  & \colhead{Secondary (B)} \\
\colhead{Parameter}  & \colhead{Component}    & \colhead{Component}
}

\startdata
a (10$^{-2}$ AU)     & 6.1305 $\pm$ 0.0084 [\em{6.205 $\pm$ 0.032}]   & 6.3179 $\pm$ 0.0092 [\em{6.322 $\pm$ 0.046}]  \\
Mass (M$_{\sun}$)    & 1.4160 $\pm$ 0.0049 [\em{1.435 $\pm$ 0.023}]   & 1.3740 $\pm$ 0.0045 [\em{1.408 $\pm$ 0.020}]  \\
\cline{2-3}
Sp Type (Barry 1970) & \multicolumn{2}{c}{F9 IVw} \\
System Distance (pc) & \multicolumn{2}{c}{36.08  $\pm$ 0.19  [\em{36.93 $\pm$ 0.56}]} \\
$\pi_{orb}$ (mas)    & \multicolumn{2}{c}{27.74  $\pm$ 0.15  [\em{27.08 $\pm$ 0.41}]} \\
Bolometric Flux (10$^{-7}$ erg cm$^{-2}$ s$^{-1}$) & \multicolumn{2}{c}{3.074 $\pm$ 0.021}\\
\cline{2-3}
T$_{eff}$ (K)  & 6130 $\pm$ 100                  & 6230 $\pm$ 150                    \\
Bolometric Flux (10$^{-7}$ erg cm$^{-2}$ s$^{-1}$)    & 1.92 $\pm$ 0.07  &  1.16 $\pm$ 0.17 \\
Model Diameter (mas) & 0.638 $\pm$ 0.025    & 0.480 $\pm$ 0.039  \\
Radius (R$_\odot$)   & 2.474 $\pm$ 0.095    & 1.86 $\pm$ 0.15                   \\
log g                & 3.802 $\pm$ 0.033    & 4.036 $\pm$ 0.070                 \\
M$_{K-{\rm CIT}}$ (mag) & 1.261 $\pm$ 0.034 [\em{1.200 $\pm$ 0.038}]   & 1.851 $\pm$ 0.034 [\em{1.818 $\pm$ 0.039}] \\
M$_{H-{\rm CIT}}$ (mag) & 1.322 $\pm$ 0.042              & 1.882 $\pm$ 0.043  \\
M$_V$ (mag)          & 2.581 $\pm$ 0.014 [\em{2.524 $\pm$ 0.052}] & 3.066 $\pm$ 0.036 [\em{3.024 $\pm$ 0.077}] \\
$V$-$K$ (mag)        & 1.293 $\pm$ 0.032                 & 1.188 $\pm$ 0.047 \\
\enddata

\end{deluxetable}

\begin{figure}[t]
%%\plotone{figures/HD_123999--F9IVw.sed1.eps}
\plotone{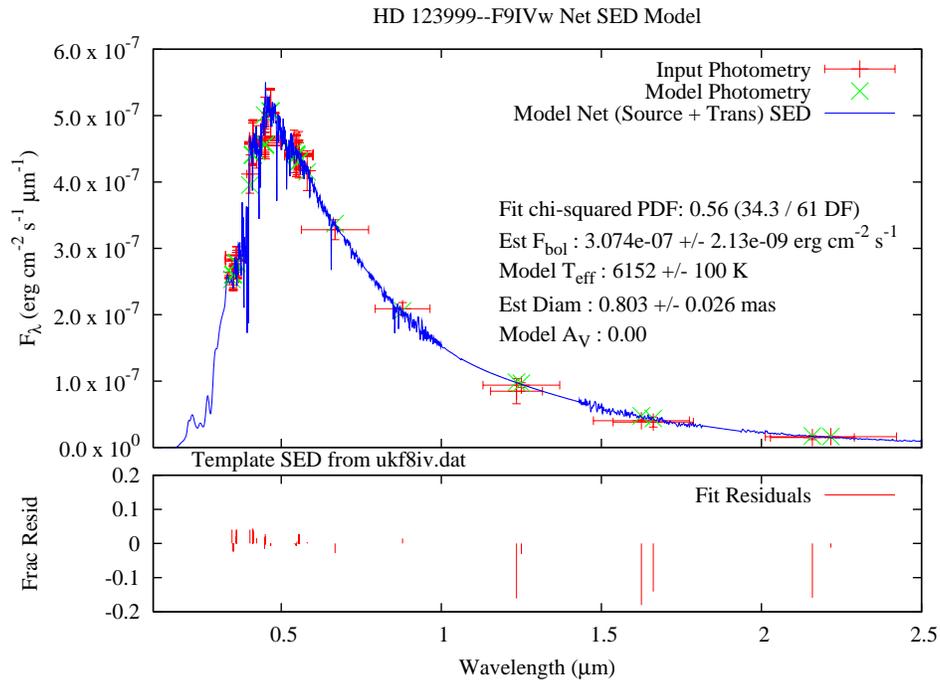}
\caption{Spectral Energy Distribution Model for 12~Boo.  Here a
template Spectral Energy Distribution (SED) template from
\citet{Pickles98} has been fit to archival photometry from SIMBAD,
2MASS, and Paper~1 to estimate the bolometric flux for 12~Boo system
(Table~\ref{tab:physics}).
%%The result (Table~\ref{tab:physics}) is in excellent agreement with
%%the published estimate of \citep{Blackwell94}.
\label{fig:12boo_SED}}
\end{figure}

\paragraph{Component Diameters, Effective Temperatures, and Radii}
The ``effective'' angular diameter of the 12~Boo system has been
estimated using the infrared flux method (IRFM) by Blackwell and
collaborators \citep{Blackwell90,Blackwell94} at approximately 0.8
mas.  At this size neither of the 12~Boo components are resolved by
PTI, and we must resort to model diameters for the component stars.
Following Blackwell, we estimate 12~Boo component diameters through
bolometric flux and effective temperature ($T_{\rm eff}$) arguments.
\citet{Blackwell94} list the bolometric flux of the 12~Boo system at
3.11$\times$10$^{-7}$ erg cm$^{-2}$ s$^{-1}$, and $T_{\rm eff}$ as
6204 K, both quoted without error estimates.  Similarly we have
analyzed archival photometry available from SIMBAD, 2MASS, and Paper~1
using an empirical model atmosphere for a solar-metallicity F8 IV star
taken from \citet{Pickles98}.  Figure \ref{fig:12boo_SED} depicts the
results of this SED modeling, resulting in a bolometric flux estimate
of 3.074 $\pm$ 0.021 $\times$10$^{-7}$ erg cm$^{-2}$ s$^{-1}$ -- in
reasonable agreement with the \citet{Blackwell94} result, and providing a
plausible error estimate.

As a check on our spectroscopic effective temperature estimates we
have made additional estimates the 12~Boo component effective
temperatures from the component colors.  The interferometric and
spectroscopic observations provide $V$ - $K$ color indices for the
components individually (Table~\ref{tab:physics}).  With these color
indices we have used effective temperature/color calibrations
published by \citet{Blackwell94} and \citet{Alonso1996} (with the
component $K$ magnitudes transformed to the Johnson system).  The
resulting component effective temperature estimates are in excellent
agreement with our adopted spectroscopic values
(\S\ref{sec:observations}, Table~\ref{tab:physics}).

Estimating the bolometric flux ratio from the observed $K$-band flux
ratio and component effective temperatures provide bolometric flux
estimates for the two components individually
(Table~\ref{tab:physics}), and these along with the effective
temperatures allow us to estimate angular diameters of 0.638 $\pm$
0.025 and 0.480 $\pm$ 0.039 mas for the primary and secondary
components respectively.  At the distance estimate to 12~Boo these
model angular diameters correspond to model component linear radii of
2.474 $\pm$ 0.095 and 1.86 $\pm$ 0.15 R$_{\sun}$ for the primary and
secondary components respectively.  Finally, coupled with our
component masses we find (log) surface gravities of 3.802 $\pm$ 0.033
and 4.036 $\pm$ 0.070 dex.  All these estimates are in good agreement
with (and more precise than) the results from Paper~1.  These linear
radii estimates are roughly a factor of two smaller than the putative
Roche lobe radii for these two stars \citep[][Eq.~1]{Iben91}, making
significant mass transfer unlikely at this stage of system evolution.

\clearpage
%% Figure 4
\clearpage

\paragraph{Component Rotation}

Tidal interaction theory predicts that in short-period binary systems
the components gravitationally interact so as to circularize the orbit
and synchronize the component rotations to the orbital period
\citep{Zahn77,Hut81}; these predictions are borne out in observation
\citep[e.g.][]{Duquennoy1991}.  The circularization and
synchronization phenomena necessarily require an energy dissipation
mechanism, generally thought to be associated with convection in the
outer envelopes of cool stars such as giants \citep{Verbunt95}.

Paper~1 noted that 12~Boo is interesting from a tidal interaction
perspective: despite the relatively short orbital period the system
orbit is modestly eccentric (Table~\ref{tab:orbit}).  The component
masses indicate both components were around F1 -- F3 at their initial
appearance on main sequence; the putative reason for the remnant
orbital eccentricity is the lack of strong convection in the
atmospheres during the components' main-sequence lives.  However, as
the 12~Boo components evolve off the main sequence their atmospheres
become much more convective, and tidal circularization and
synchronization should begin.  Once the component atmospheres become
fully convective the timescale for rotation synchronization will be
short ($\sim$ 1 Myr for the primary; Paper~1).

Several recent measurements of the rotation \vsini~of 12~Boo
components exist, offering the possibility to assess whether the two
components are currently synchronously rotating.  These 12~Boo
component rotation measurements are summarized
Table~\ref{tab:rotation}.  As in Paper~1, the consensus remains that
both 12~Boo components appear to be rotating consistent with
synchronous rates.  Within the errors and the scatter of
the measurements the secondary $v \sin i$ is also consistent with the
pseudosynchronous rate (synchronous with the orbital motion at
periastron; Hut 1981), while the primary appears to be rotating
marginally slower than the pseudosynchronous rate.

\begin{table}
\begin{center}
\begin{small}
\begin{tabular}{ccc}
\hline
		 	& Primary           & Secondary \\
	         	& \vsini            & \vsini    \\
			& (km s$^{-1}$)     & (km s$^{-1}$) \\
\hline
\cite{Balachandran90}   & 10 $\pm$ 3	    & 		\\
\cite{DeMedeiros97}	& 12.7 $\pm$ 1      &		\\
DU99			& 12.5 ($\pm$ 1)    & 9.5 ($\pm$ 1)  \\
Paper~1                 & 13.1 $\pm$ 0.3    & 10.4 $\pm$ 0.3 \\
\cite{Shorlin2002}      & 14.0 $\pm$ 3.0    & 12.0 $\pm$ 3.0 \\
\cite{Reiners2003}      & 15.0 $\pm$ 1.0    &                \\
This work               & 14.0 $\pm$ 1.0    & 12.0 $\pm$ 1.0 \\
%%\hline
%%Composite 	        & 13.04 $\pm$ 0.26  & 10.45 $\pm$ 0.28 \\
\hline
Model Synch Rotation		& 12.4 ($\pm$ 1.1)  & 9.3 ($\pm$ 0.8) \\
Model Pseudo-Synch Rotation	& 15.2 ($\pm$ 1.4)  & 11.4 ($\pm$ 1.0) \\
\hline
\end{tabular}
\end{small}
\caption{\vsini~Measurements for 12~Boo Components.  Summarized here
are recent \vsini~measurements for the 12~Boo system components,
including this work.  For references where a single \vsini~measurement
is listed we have assumed this pertains to the primary component.
DU99 does not list errors for their component \vsini~estimates; we
have arbitrarily taken 1 km s$^{-1}$ so as to be consistent with the
characteristic accuracies of earlier CORAVEL determinations \citep[see
discussions in][]{DeMedeiros96,DeMedeiros97}.  For comparison we give
model estimates of \vsini~for synchronous and pseudosynchronous
rotation of the two components given the physical sizes discussed in
\S~\ref{sec:physics}.  Both 12~Boo components would appear to be
rotating at rates consistent with synchronous rotation.
\label{tab:rotation}}
\end{center}
\end{table}

\section{Comparisons With Stellar Models}
\label{sec:models}
With our estimates of the component masses, absolute magnitudes, color
indices, and effective temperatures derived from our measurements and
orbital solution (Table~\ref{tab:physics}), we proceed in this section
to examine the 12~Boo components in the context of recent stellar
evolution models. 

\clearpage
%% Figure 5
\begin{figure}[p]
\epsscale{1.1}
%%\plottwo{figures/12boo.MKisochrone.eps}{figures/12boo.MKCMD.eps}\\
%%\plottwo{figures/12boo.MVisochrone.eps}{figures/12boo.MVCMD.eps}\\
\plottwo{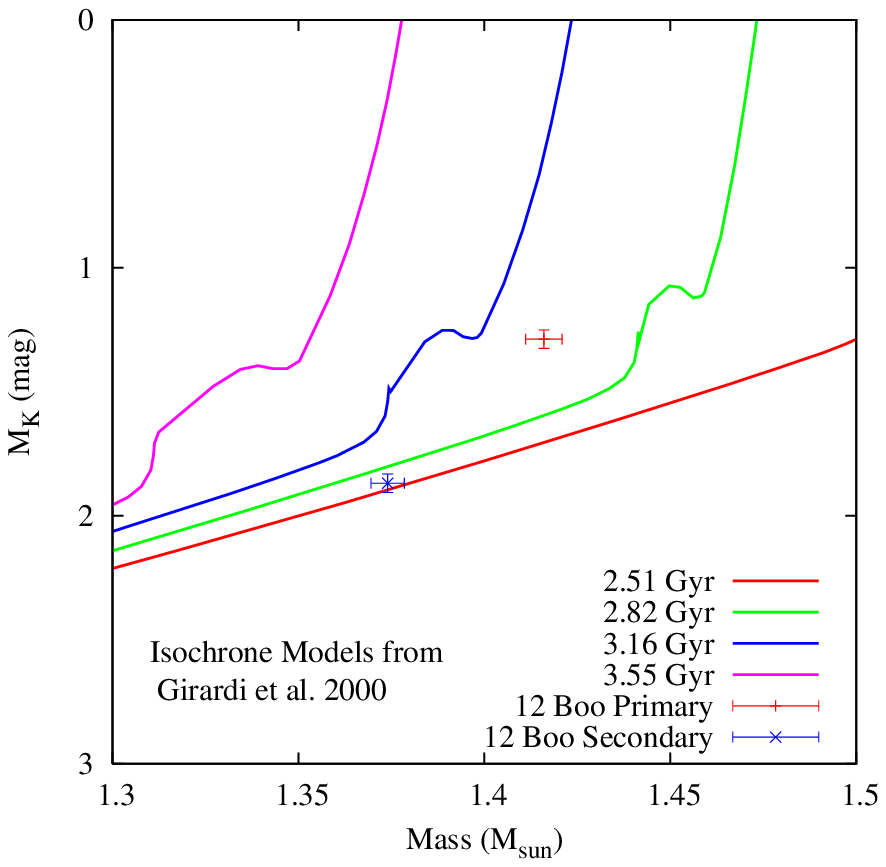}{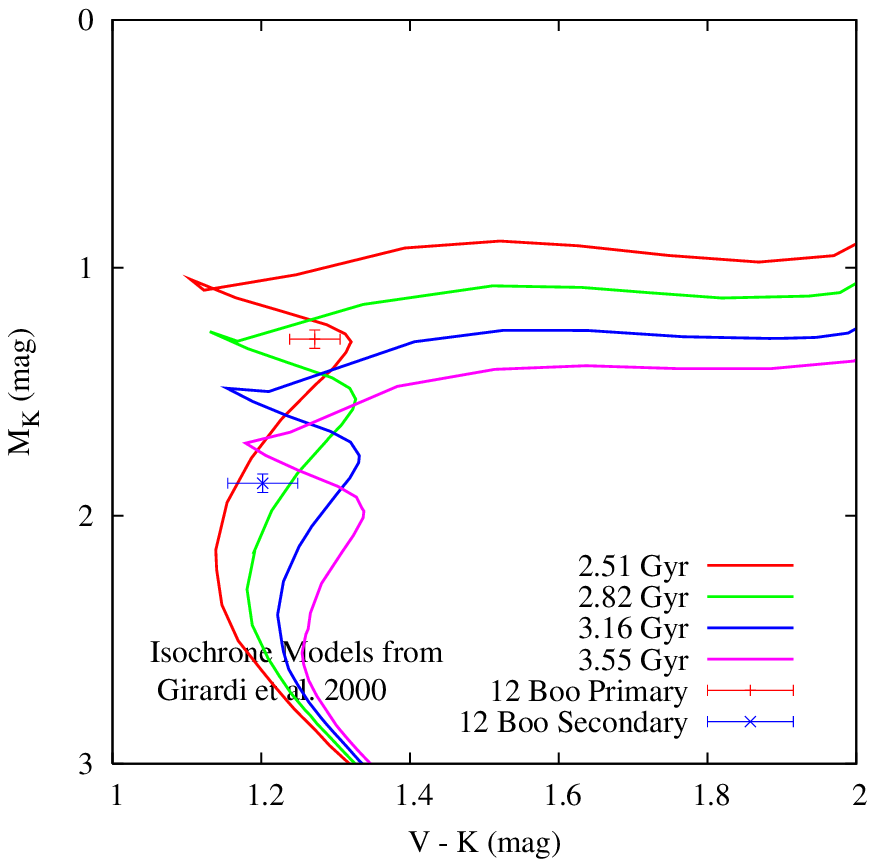}\\
\plottwo{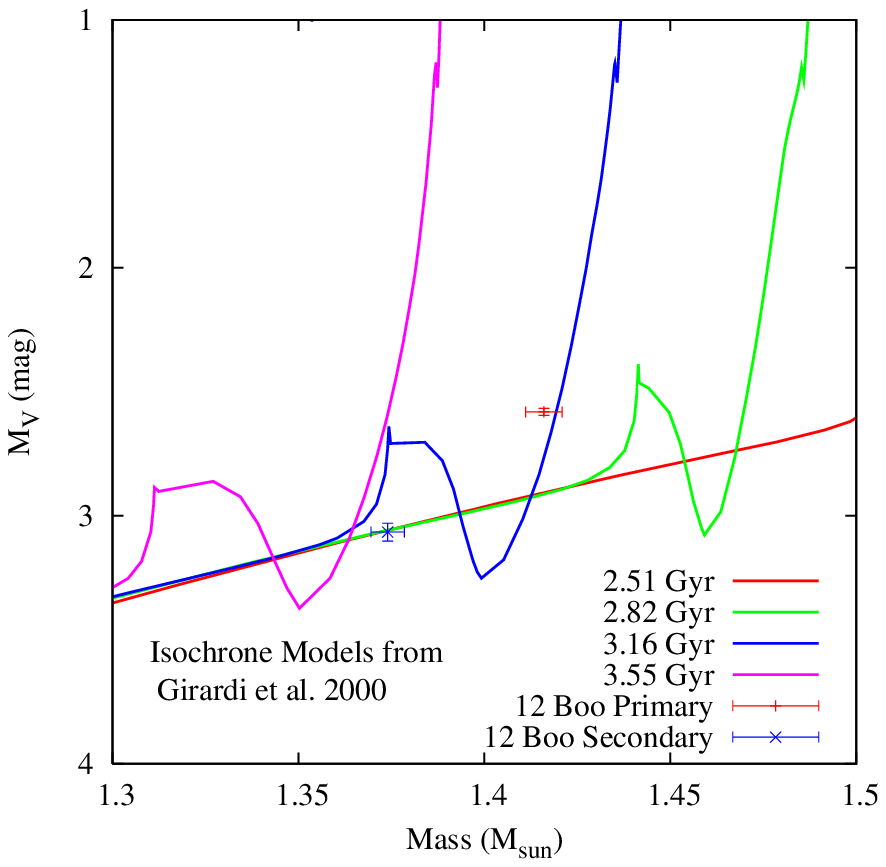}{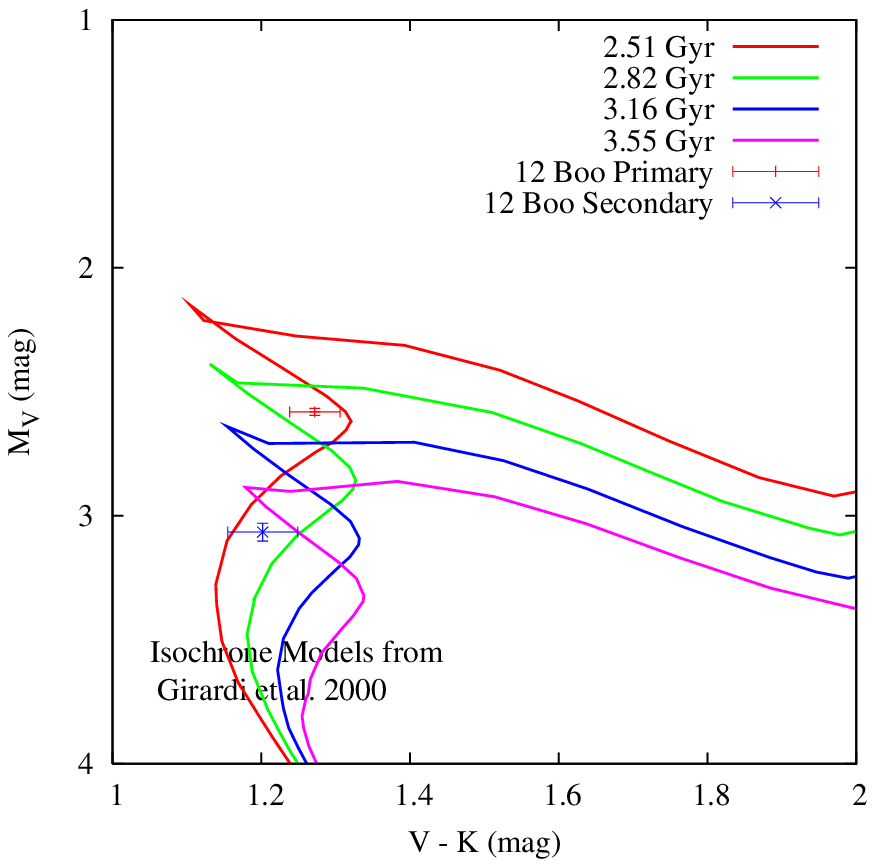}\\
\caption{Comparison of the 12~Boo Component Parameters with Girardi et
al.~2000 (G2000) Stellar Models.  Comparisons between our observed
12~Boo component parameters and G2000 models are shown in observable
mass-magnitude and color-magnitude spaces.
\label{fig:12boo_Girardi2000}}
\end{figure}
\clearpage

In Paper~1 we compared the measurements for the 12~Boo components with
models from the Padova series by \citet{Bertelli1994}. Since then the
input physics of these particular models has been updated mainly by
incorporating improvements in the equation of state and in the
opacities, as described by \citet[hereafter G2000]{Girardi2000}. In
Figure~\ref{fig:12boo_Girardi2000} the observed properties of the
12~Boo components are shown against four isochrones for solar
metallicity from G2000, in various planes.  The panels on the left
show the $V$ and $K$ absolute magnitudes versus mass (diagrams for $H$
are similar and are not shown here).  \citep[For purposes of model
comparisons the component infrared magnitudes have been transformed
from the CIT to the Johnson system using color conversions
from][]{Bessell1988}.  No single isochrone appears to fit the
observations within the error bars. The diagrams on the right suggest
that an isochrone between 2.5 Gyr and 2.8 Gyr might provide a good fit
in the color-magnitude plane, with both stars located near the end of
their hydrogen-burning phase.  Paper~1 reached a similar conclusion on
the system age estimate based on Padova models.  However, as indicated
by the left-hand figures the model masses for such an age would not
agree with the measured component parameters.

\clearpage
%% Figure 6
\begin{figure}[p]
\epsscale{1.1}
%%\plottwo{figures/fig_mk_vmk.ps}{figures/fig_mk_teff.ps}\\
%%\plottwo{figures/fig_mv_vmk.ps}{figures/fig_mv_teff.ps}\\
\plottwo{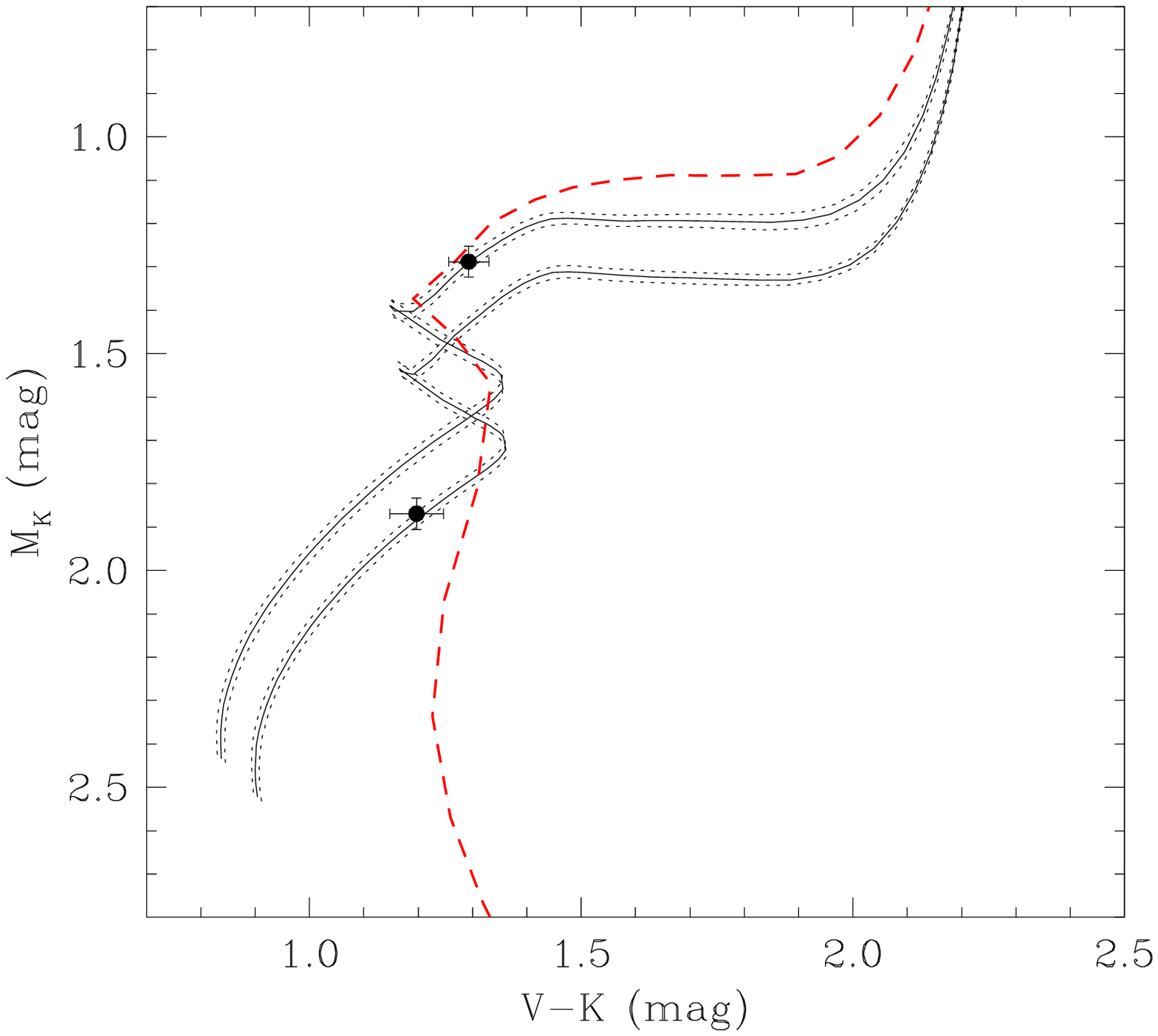}{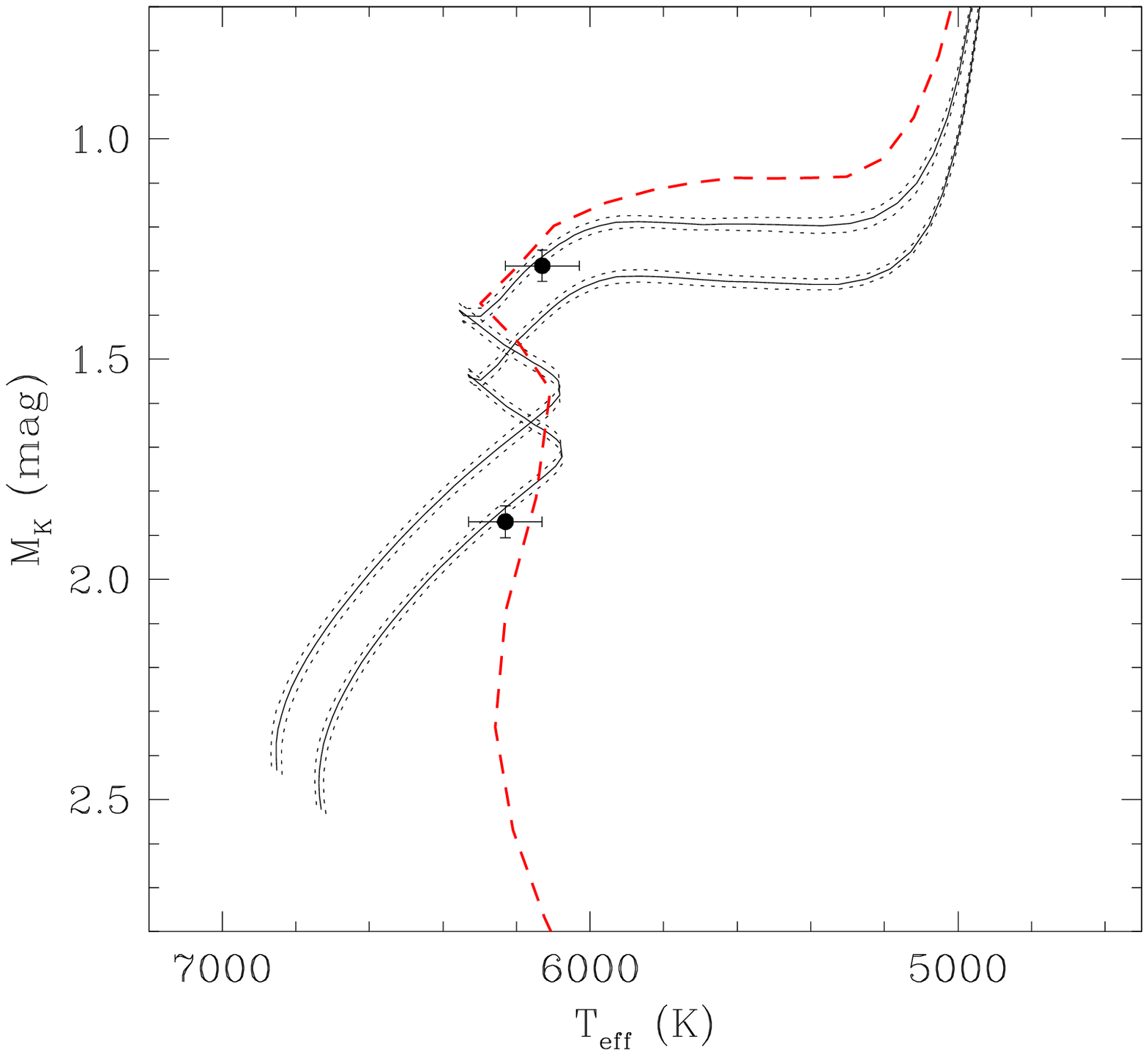}\\
\plottwo{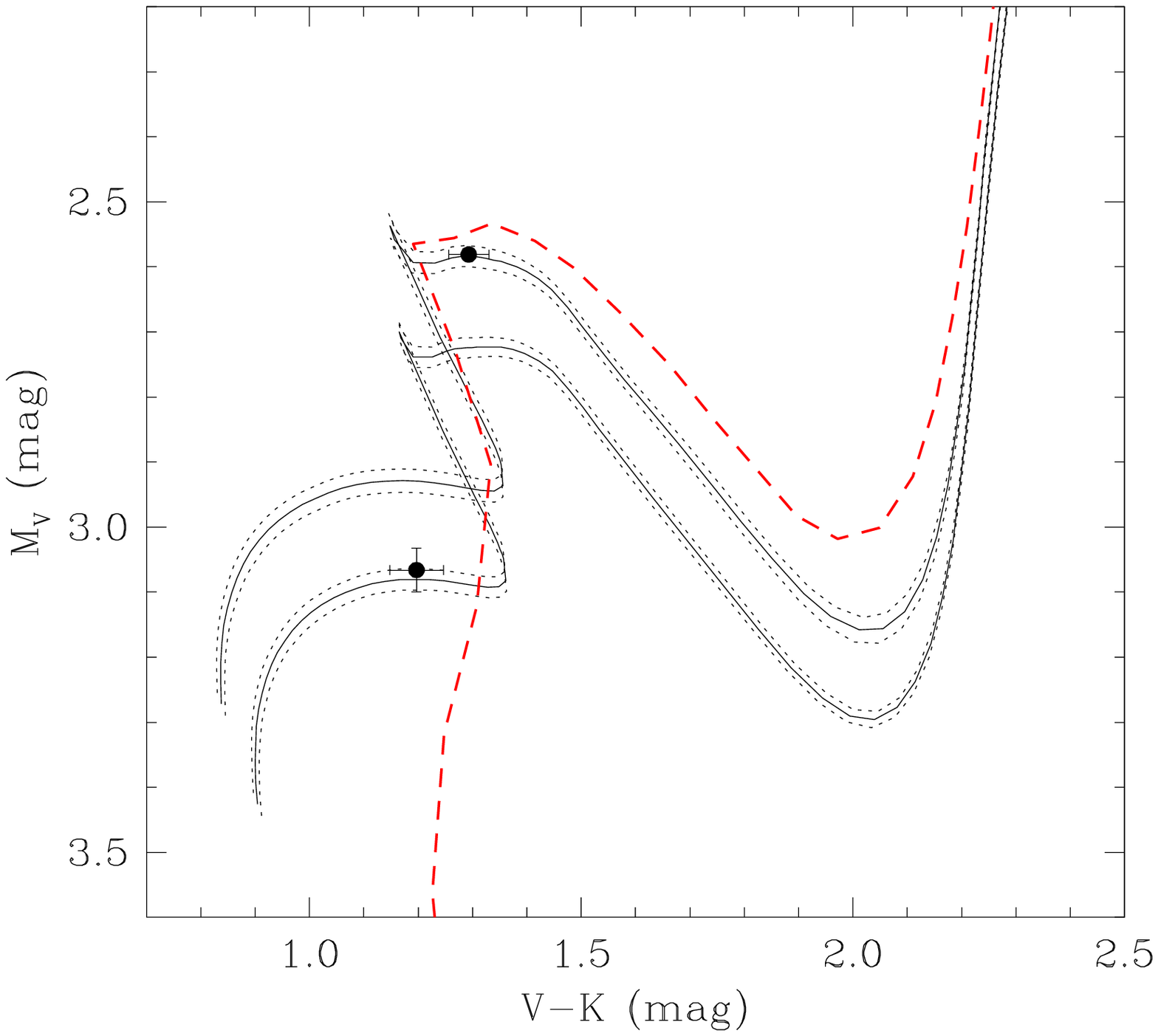}{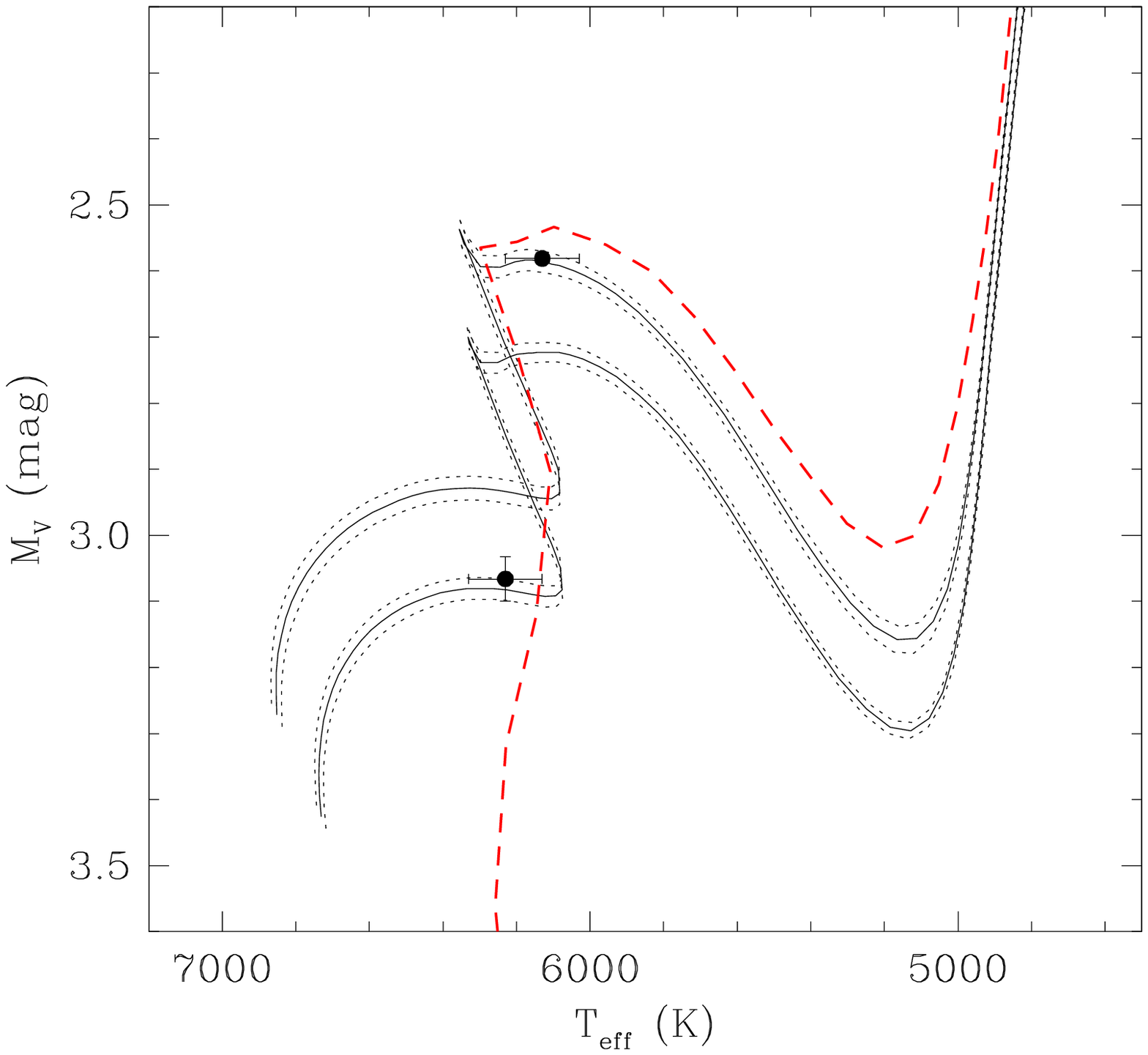}\\
\caption{Comparison of the 12~Boo Component Parameters with
Yonsei-Yale (Y$^2$) Stellar Models.  Comparisons between our observed
12~Boo component parameters and Y$^2$ models are shown in
color-magnitude and temperature-magnitude spaces.  The figures give
mass tracks computed for the particular masses of the 12~Boo
components (shown in black including mass uncertainties), and the
best-fit single isochrone to the parameters of both components (age of
3.2 Gys -- shown in the red dashed line).
\label{fig:12boo_yale2004}}
\end{figure}
\clearpage

\begin{figure}[p]
\epsscale{1.1}
%%\plottwo{figures/fig_r_vmk.ps}{figures/fig_logg_vmk.ps}\\
\plottwo{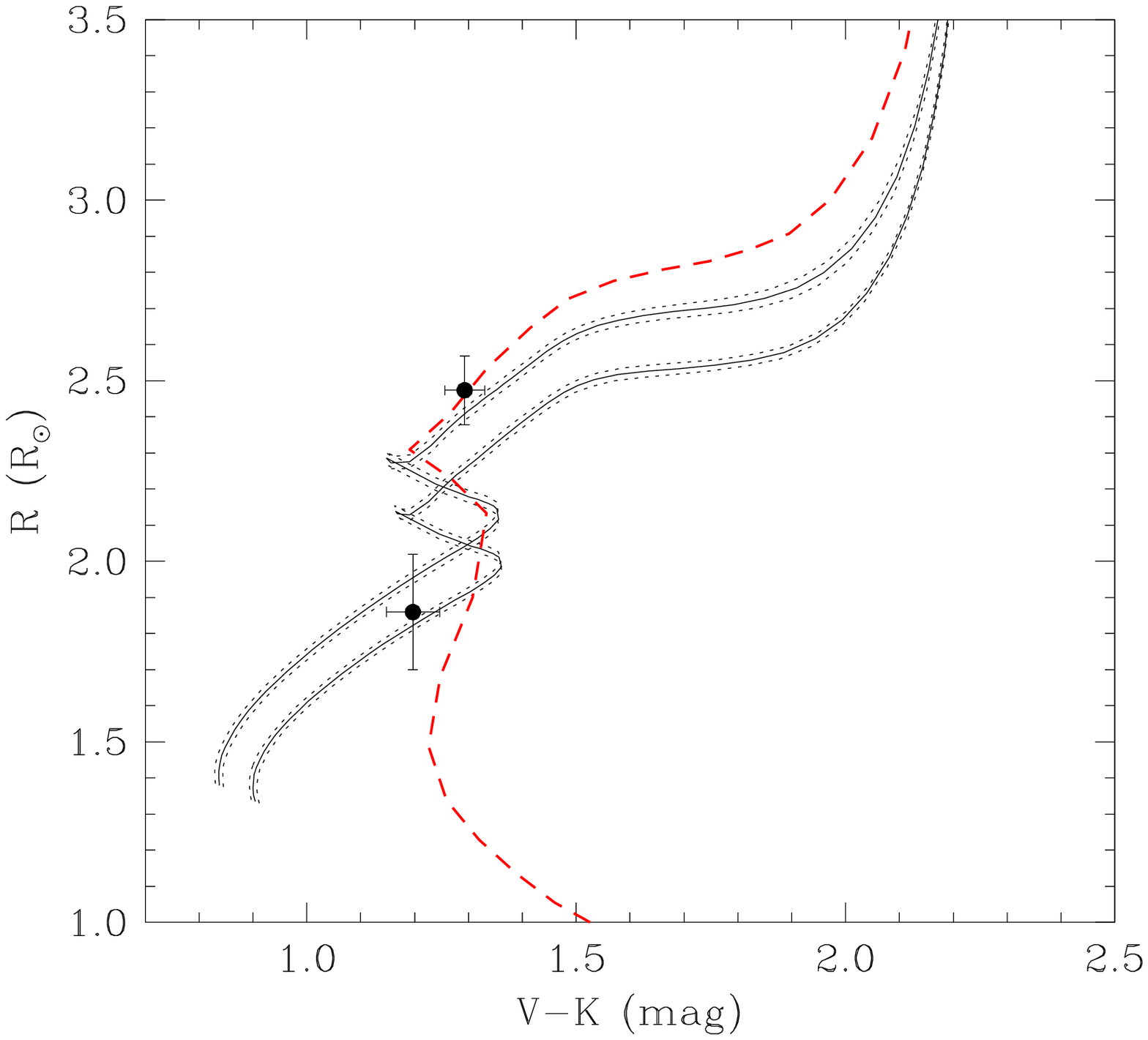}{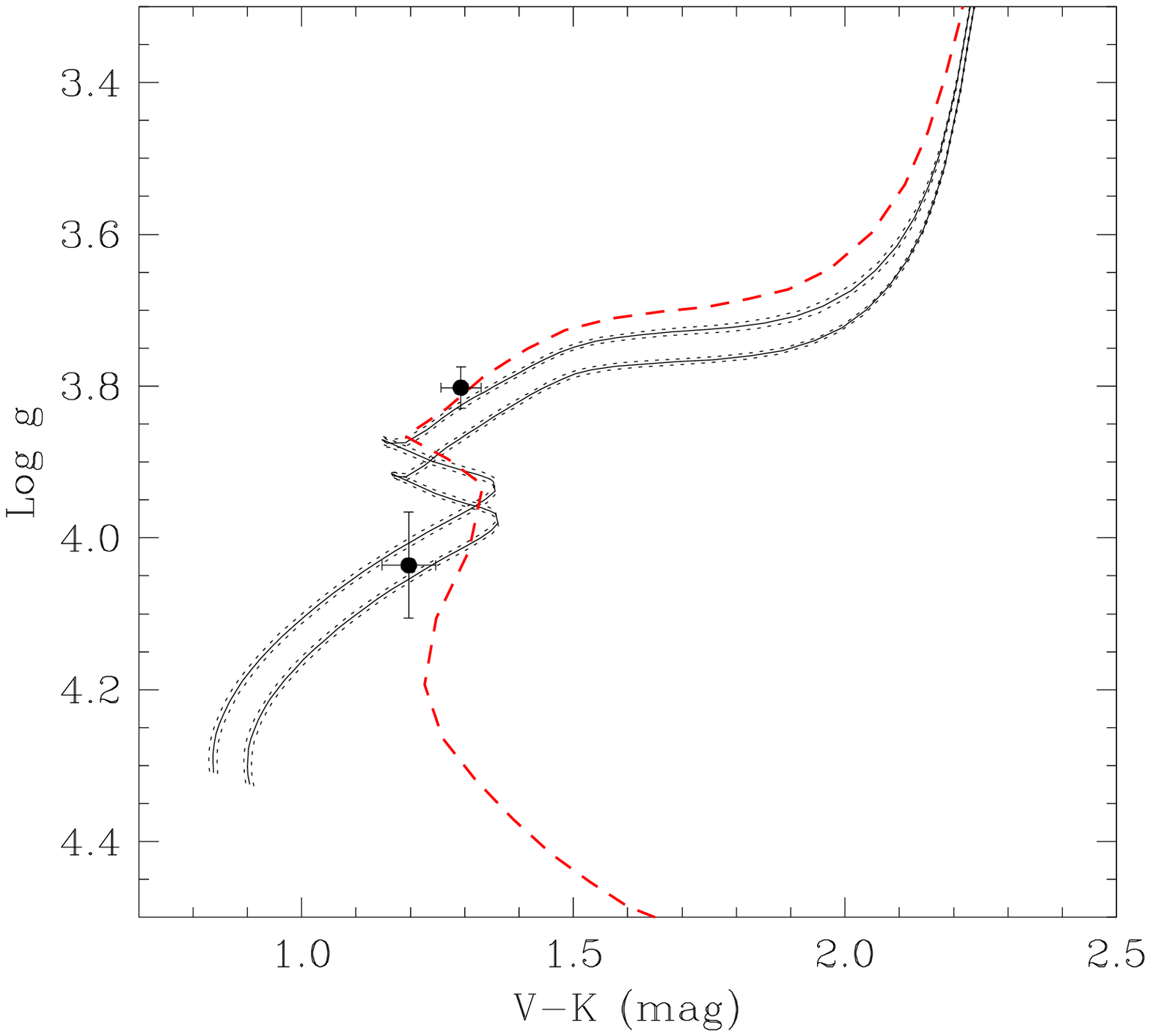}\\
\caption{Comparison of the 12~Boo Component Radii with Y$^2$ Models.
Comparisons of the estimated component radii and surface gravity with
predictions from the Y$^2$ models.  Mass tracks for the two system
components are given in solid black lines, with the 3.2 Gyr isochrone
is given in the red dashed line.
\label{fig:logg_yale2004}}
\end{figure}

Figure~\ref{fig:12boo_yale2004} compares the observed quantities to
models from the Yonsei-Yale (Y$^2$) collaboration \citep{Yi2001,
Demarque2004}. The Y$^2$ models use similar physics to G2000, but
differ in a number of details including the radiative opacities, the
equation of state, the treatment of convective core overshooting, and
the implementation of helium and heavy element diffusion (not
accounted for in G2000).  In this case we show evolutionary tracks for
the exact component masses determined from the physical orbit
(Table~\ref{tab:physics}), using an interpolation routine described by
\cite{Yi2003}.  Absolute magnitudes in $V$ and $K$ are displayed
against both $V-K$ and effective temperature as estimated from our
spectra, so that each panel displays the constraint from three
observables at the same time (shown with their uncertainties) rather
than two as in the previous figure.  The one-sigma mass uncertainties
are indicated by the dotted lines bracketing the tracks.  To show the
constraint on coevality, which we assume to hold for this binary, an
isochrone from the same series of models is also represented in each
panel. An age of 3.2 Gyr provides the best overall fit to the
observations; this is significantly older than the system age estimate
from Paper~1 based on Bertelli models.  As mentioned in
\S~\ref{sec:physics}, the metallicity determinations in the literature
suggest a near-solar composition for 12~Boo.  The Y$^2$ models seem to
agree with that assessment; in surveying a range of metallicities
allowed by previous studies we found the best agreement between our
observational parameters and the model predictions at solar abundance
($+0.0$ dex/Z=0.01812 in Y$^2$ models).  While our estimates of the
surface gravities and absolute radii for the stars rely not on the
physical orbit but on other radiative properties, they do enter weakly
into the orbital solution (through the component angular diameters) as
well as our spectroscopic estimate of the effective temperatures
(\S~\ref{sec:observations}).  The comparison of our inferred $\log g$
and radius (Table~\ref{tab:physics}) values with the \cite{Yi2003}
models is shown in Figure~\ref{fig:logg_yale2004}.

%%%INSERT FIGURES OF LOG G AND R VERSUS TEFF (OR V-K)  ;-)
%% Figure 7

\clearpage
\clearpage

\begin{figure}[p]
\epsscale{0.7}
%%\plotone{figures/overshoot_yale2004.ps}\\
\plotone{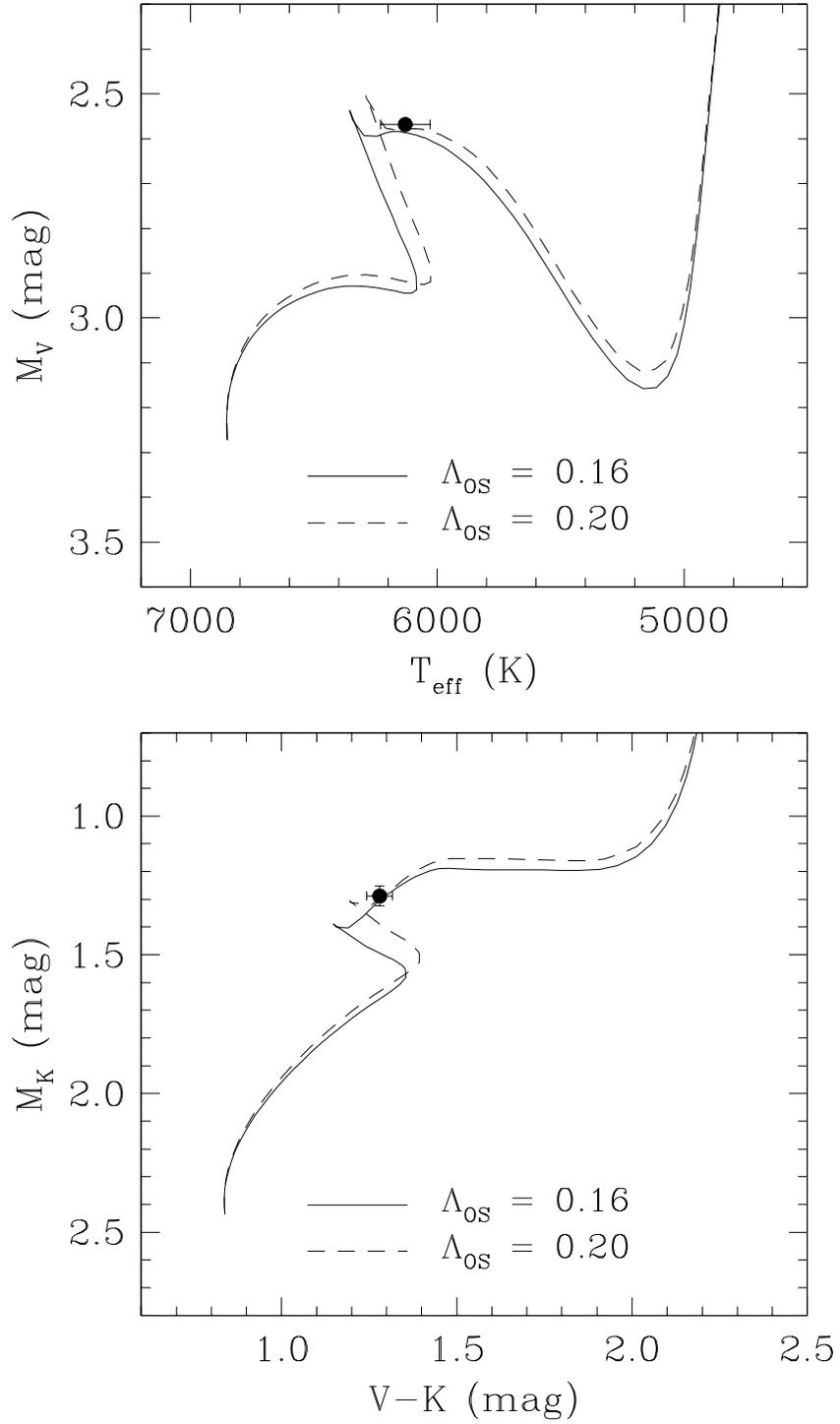}\\
\caption{Effects of Different Convective Overshooting Parameters on the Y$^2$ Models
\label{fig:overshoot_yale2004}}
\end{figure}

Figure~\ref{fig:12boo_yale2004} suggests the secondary of 12~Boo is
comfortably in the main-sequence stage, while the primary would appear
to be near the beginning of the rapid phase of evolution where it
burns hydrogen in a shell -- the so-called Hertzsprung gap.  The
duration of this phase is only about 4\% of the main-sequence lifetime
for a star of this mass.  Although it is a priori unlikely that we
would find a star in this state, the possibility can certainly not be
excluded
\citep[e.g.~see][]{Andersen1990,Fekel2001,Parsons2004}. However, we
note that a minor increase in the amount of convective core
overshooting (a free parameter in the models) could easily extend the
main sequence enough to bring agreement with the observations for the
primary star, placing it at the end of the hydrogen-burning phase
rather than in the Hertzsprung gap. The treatment of overshooting in
these models follows the recent prescription described by
\cite{Demarque2004}, in which the overshooting parameter $\Lambda_{\rm
OS}$ (in units of the pressure scale height $H_p$) increases gradually
from 0.00 to 0.20 as a function of mass in the regime in which a
convective core develops ($M_{crit}^{conv} \sim 1.2$~M$_{\sun}$ for
solar composition). This was deemed a more realistic approximation
than that adopted in the previous release of the Y$^2$ models
\citep{Yi2001}, in which $\Lambda_{\rm OS} = 0.00$ for masses up to
$M_{crit}^{conv}$ and $\Lambda_{\rm OS} = 0.20$ for larger masses.
Given the mass of the 12~Boo primary (Table~\ref{tab:physics}), the
interpolation performed to produce the track for
Figure~\ref{fig:12boo_yale2004} results in an effective overshooting
parameter of approximately 0.16.  With the old prescription the
overshooting would be 0.20. In Figure~\ref{fig:overshoot_yale2004} we
illustrate the effect of changing $\Lambda_{\rm OS}$ by showing the
primary tracks for both values, where the solid curve is the same
model as in Figure~\ref{fig:12boo_yale2004} (new overshooting
prescription). As expected the track corresponding to $\Lambda_{\rm
OS} = 0.20$ (old prescription) has a more extended main sequence
reaching slightly cooler temperatures, and comes closer to matching
the observed location of 12~Boo~A at the very end of the core burning
phase.  While both sets of model predictions are consistent with the
observed properties of the 12~Boo primary, it seems a priori more
likely that the star is at the end of its main-sequence life.  If the
12~Boo primary is at the end of main sequence, the observations would
seem to place fairly tight constraints on overshooting that suggest
somewhat larger values of $\Lambda_{\rm OS}$ than adopted by
\cite{Demarque2004}.

\clearpage
%% Figure 8
\clearpage

\begin{figure}[p]
\epsscale{0.6}
%%\plotone{figures/compare_iso.ps}\\
\plotone{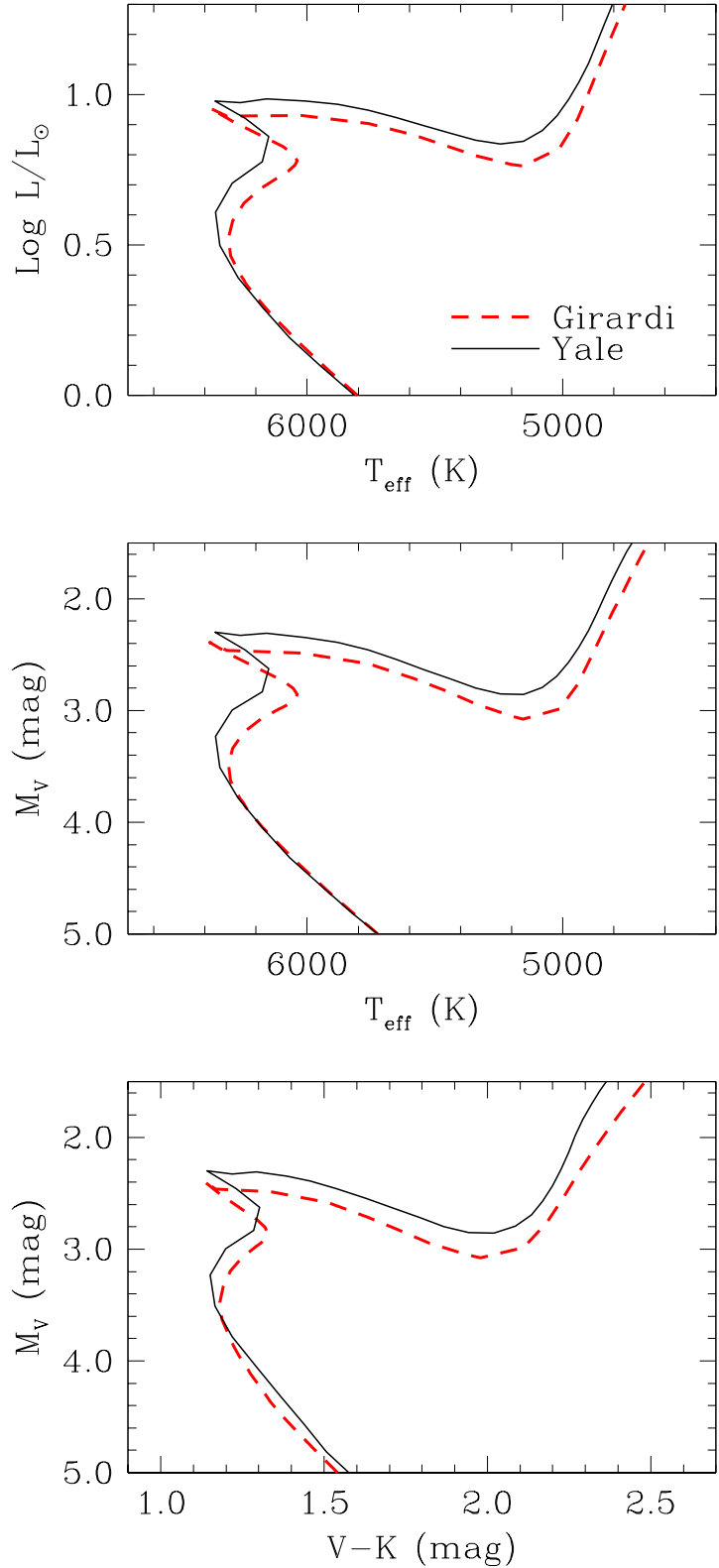}\\
\caption{Direct Comparisons of G2000 and Y$^2$ Models in Theoretical
and Observational Color-Magnitude Spaces.  The three panels show
comparisons of 2.82 Gyr solar metallicity isochrones near the
main-sequence end from G2000 (red) and Y$^2$ (black) in theoretical
(Luminosity vs. T$_{eff}$ -- top), semi-observational (M$_V$
vs. T$_{eff}$ -- middle), and observational (M$_V$ vs. $V-K$ --
bottom) color-magnitude spaces.
\label{fig:compare_iso}}
\end{figure}

The apparent agreement between Y$^2$ model predictions and our
measurements is in contrast to the apparent disagreement of our
results with the Padova/G2000 model series.  The differences between
the Y$^2$ and Padova models are more directly appreciated in
Figure~\ref{fig:compare_iso}, where we show 2.82 Gyr ($\log age =
9.45$) isochrones from both series for the same (solar) metallicity.
From top to bottom we depict the isochrones from the purely
theoretical plane (luminosity versus $T_{eff}$) to the purely
observational plane (absolute magnitude versus $V-K$ color).  While
there is excellent agreement for unevolved stars in the top two
diagrams (lower main sequence), the end of the main-sequence region
highlights the subtle differences that have to do with the details in
the input physics. In the lower panel the discrepancies extend also to
the unevolved stars; this is due to differences in the
color-temperature calibrations between the models.  Over the
temperature range shown in the figure the Padova isochrone relies on
color and bolometric correction tables based on theoretical model
atmospheres \citep[see][]{Bertelli1994}, while the Y$^2$ model relies
on semi-empirical tables by \cite{Lejeune1998}.
	
\clearpage
%%FIGURE compare_iso.ps
%% Figure 9
\clearpage

\section{Summary and Discussion}
\label{sec:summary}
By virtue of our interferometric resolution and the precision of the
radial velocity data we are able to determine an accurate physical
orbit for 12~Boo, resulting in accurate physical parameters for the
12~Boo constituents and an accurate system distance.  Our 12~Boo
distance estimate is in excellent agreement with the Hipparcos
determination.  Our finding of unexpectedly large relative $K$, $H$,
and $V$-magnitude differences in the two nearly-equal mass 12~Boo
components is understood in the context that the system is in a unique
evolutionary state, with the primary component apparently making its
transition off the main sequence.  12~Boo component rotation
measurements are consistent with synchronous rotation for the system
components, and at least the primary is less consistent with
pseudosynchronous rotation.

The results of comparing the 12~Boo components with stellar models are
mixed.  While we see relatively good agreement between the component
physical parameters and the Y$^2$ evolutionary (mass) tracks, the
agreement with the G2000 isochrones is not nearly as good.  Further,
the discrepancy seems to be intrinsic; Figure~\ref{fig:compare_iso}
illustrates that fundamental differences exist between the Y$^2$ and
G2000 models in both theoretical and observational spaces near the end
of the main sequence.  Our measured 12~Boo component parameters are
clearly in much better agreement with the Y$^2$ model predictions for
intermediate-mass stars near the end of the main sequence.

\begin{figure}[p]
\epsscale{1.1}
%%\plottwo{figures/mk_vmkAge1.eps}{figures/mk_vmkAge2.eps}\\
%%\plottwo{figures/mv_vmkAge1.eps}{figures/mv_vmkAge2.eps}\\
\plottwo{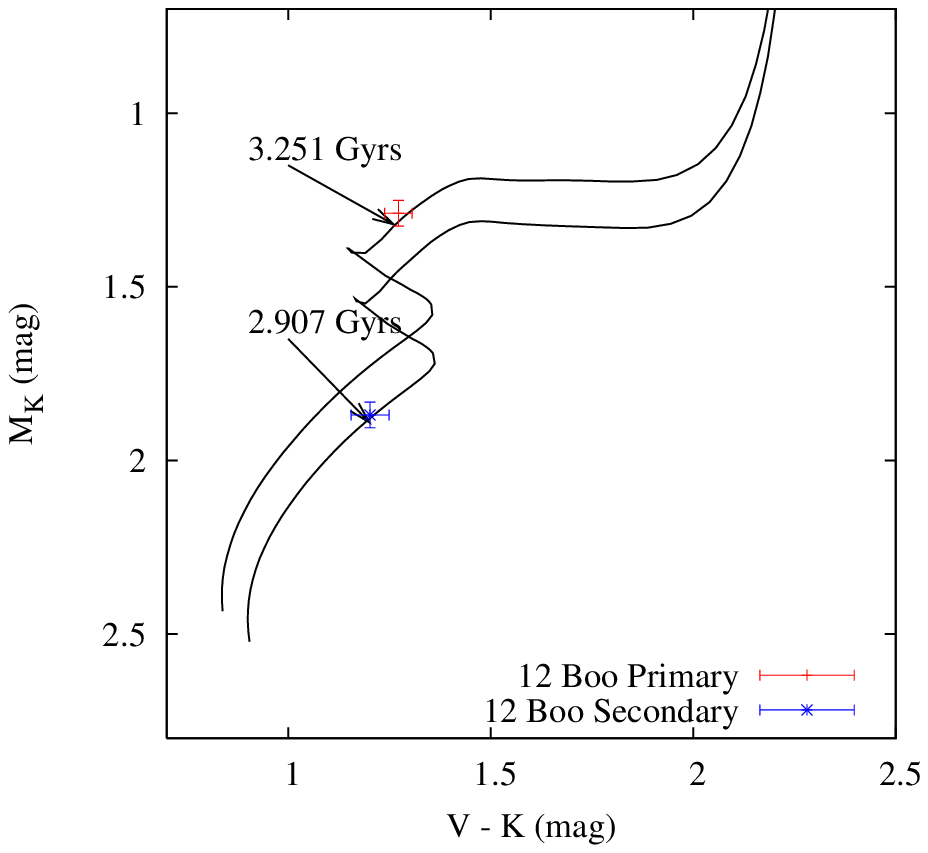}{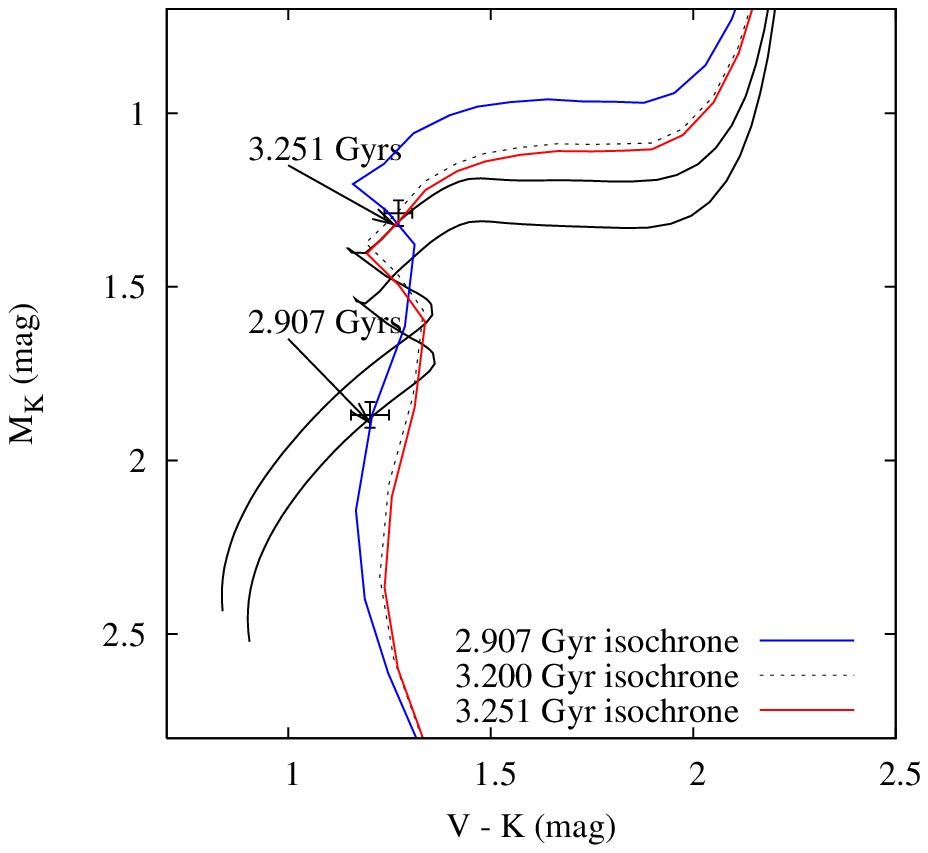}\\
\plottwo{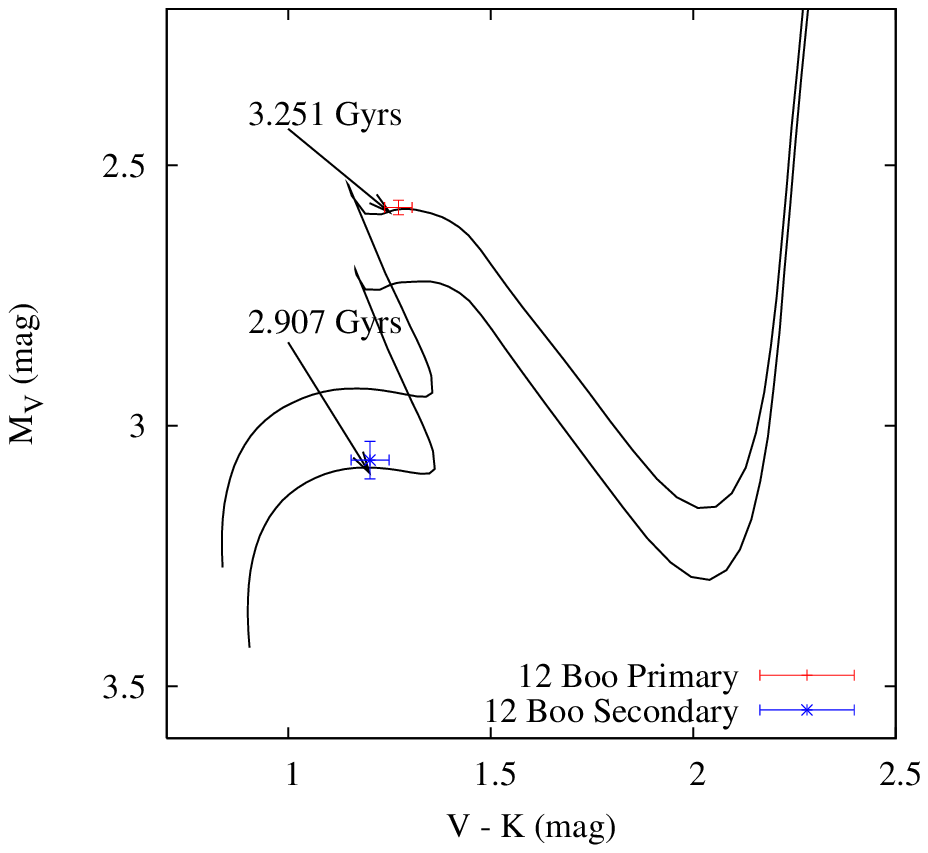}{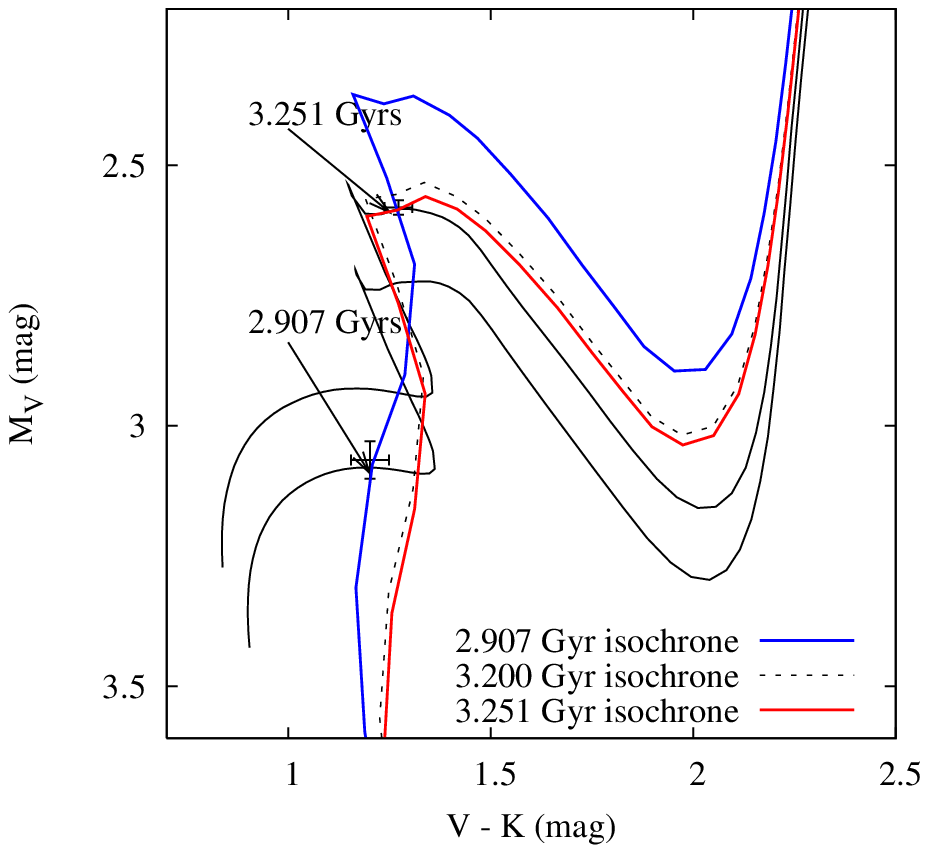}\\
\caption{12~Boo Component Age Estimates from Y$^2$ Stellar Models.
The left panels indicate best-match component ages estimated from
Y$^2$ mass tracks in observational $M_K$ and $M_V$ vs $V - K$ planes.
The right panels show the same quantities, superimposing Y$^2$
isochrones computed for these best-match component age estimates.
Presumably the apparent 340 My component age discrepancy cannot be
physical.
\label{fig:componentAges}}
\end{figure}

Further, it is interesting to note that the agreement between our
observations and the Y$^2$ mass tracks is significantly better than
the agreement with the best-fit Y$^2$ isochrone at 3.2 Gyr.
Presumably the 12~Boo components must be coeval, so the larger
mismatch in the isochrone prediction must be indicative of a remaining
discrepancy between the observations and Y$^2$ models.  This
discrepancy is illustrated in Figure~\ref{fig:componentAges}, which
depicts the 12~Boo components and their mass tracks in the same
observational color-absolute magnitude spaces given in
Figure~\ref{fig:12boo_yale2004}.  The left panels in
Figure~\ref{fig:componentAges} focus on the observed component
parameters and Y$^2$ mass tracks, and in particular indicate the ages
on the mass tracks that best match the component parameters.  We find
the Y$^2$ tracks would indicate best-match ages of 3.25 and 2.91 Gyrs
for the primary and secondary components respectively.  The right
panels in Figure~\ref{fig:componentAges} show these same spaces with
Y$^2$ isochrones computed for the specific best-match component ages.
Presumably this apparent discrepancy in the component ages cannot be
physical.  It seems likely that the unique evolutionary state of
12~Boo could provide important observational constraints on the
physical evolution of intermediate-mass stars making their transition
off the main sequence.

\clearpage
%% Figure 10
\clearpage

Finally, Figures~\ref{fig:12boo_yale2004}, \ref{fig:logg_yale2004},
and \ref{fig:componentAges} lead to the tempting inference that the
12~Boo primary is early in its first transition across the Hertzsprung
gap to the base of the red giant branch.  However, as discussed in
\S~\ref{sec:models}, Figure~\ref{fig:overshoot_yale2004} shows that at
the apparent evolutionary state of the 12~Boo primary relatively small
changes in the physics of the stellar models can make significant
changes in the model predictions.  Particularly coupled with the {\em
apparent} age discrepancy in the 12~Boo components indicated by the
models, caution suggests that a Hertzsprung-gap interpretation for the
12~Boo primary should be provisional only.

\acknowledgements 

Work done with the Palomar Testbed Interferometer was performed at the
Michelson Science Center, California Institute of Technology under
contract with the National Aeronautics and Space Administration.
Interferometer data were obtained at Palomar Observatory using the
NASA Palomar Testbed Interferometer, supported by NASA contracts to
the Jet Propulsion Laboratory.  Science operations with PTI are
conducted through the efforts of the PTI Collaboration
(http://huey.jpl.nasa.gov/palomar/ptimembers.html), and we acknowledge
the invaluable contributions of our PTI colleagues.  We particularly
thank Kevin Rykoski for his professional operation of PTI.

We thank Joe Caruso, Bob Davis, David Latham, Robert Stefanik, and Joe
Zajac for obtaining many of the spectroscopic observations used here.
GT acknowledges partial support from NASA's MASSIF SIM Key Project
(BLF57-04) and NSF grant AST-0406183.

Work done with the NPOI interferometer was performed through a
collaboration between the Naval Research Lab and the US Naval
Observatory in association with Lowell Observatory, and was funded by
the Office of Naval Research and the Oceanographer of the Navy.

This research has made use of the SIMBAD database, operated at CDS,
Strasbourg, France; NASA's Astrophysics Data System Abstract Service;
and services from the Michelson Science Center, California Institute
of Technology, http://msc.caltech.edu.


\begin{thebibliography}{99}



\bibitem[Abt \& Levy(1976)]{Abt76}
Abt, H.~ and Levy, S.~1976, \apjs~30, 273.

\bibitem[Alonso et al.(1996)]{Alonso1996}
Alonso, A., Arribas, S., and Martinez-Roger, C.~1996, \aap 313, 873.

\bibitem[Alonso et al.(1999)]{Alonso1999}
Alonso, A., Arribas, S., and Martinez-Roger, C.~1999, \aaps 140, 261.

\bibitem[Andersen et al(1990)]{Andersen1990}
 Andersen, J., Nordstr\"om, B., \& Clausen, J.\ V. 1990, \apj, 363,
L33

\bibitem[Armstrong et al.(1992)]{Armstrong92b}
Armstrong, J.T.~et al.~1992, \aj~104, 2217.

\bibitem[Armstrong et al.(1998)]{Armstrong1998}
Armstrong, J.T.~et al.~1998, \apj~496, 550.

\bibitem[Balachandran(1990)]{Balachandran90}
Balachandran, S.~1990, \apj~354, 310.

\bibitem[Barry(1970)]{Barry70}
Barry, D.~1970, \apjs~19, 281.

\bibitem[Bertelli et al.(1994)]{Bertelli1994}
Bertelli, G., Bressan, A., Chiosi, C., Fagotto, F., and Nasi, E.~1994 (B94), \aaps~106, 275.

\bibitem[Bessell \& Brett(1988)]{Bessell1988}
Bessell, M.~and Brett, J.~1988, \pasp~100, 1134.

\bibitem[Blackwell et al.(1990)]{Blackwell90}
Blackwell, D., Petford, A., Arribas, S., Haddock, D., and Selby, M.~1990, \aaps~232, 396.

\bibitem[Blackwell \& Lynas-Gray(1994)]{Blackwell94}
Blackwell, D., and Lynas-Gray, A.~1994, \aap~282, 899.

\bibitem[Boden et al.(1998)]{Boden98}
Boden, A.F.~et al.~1998, \procspie~3350, 872.

\bibitem[Boden et al.(1999a)]{Boden99a}
Boden, A.F.~et al.~1999a, \apj~515, 356.

\bibitem[Boden et al.(1999b)]{Boden99b}
Boden, A.F.~et al.~1999b, \apj~527, 360.

\bibitem[Boden et al.(2000)]{Boden2000}
Boden, A.F., Creech-Eakman, M., and Queloz, D.~2000, \apj 536, 880 (Paper~1).

\bibitem[Boden \& Lane(2001)]{BL2001}
Boden, A.F.~and Lane, B.F.~2001, \apj~547, 1071.

\bibitem[Campbell \& Wright(1900)]{Campbell1900}
Campbell, W.W.~and Wright, W.H.~1900, \apj~12, 254.

\bibitem[Colavita et al.(1999)]{Colavita99a}
Colavita, M.M.~et al.~1999, \apj~510, 505 (astro-ph/9810262).

\bibitem[Colavita(1999)]{Colavita99b}
Colavita, M.~1999, \pasp~111, 111 (astro-ph/9810462).

\bibitem[Demarque et al.(2004)]{Demarque2004}
 Demarque, P., Woo, J.-H., Kin, Y.-C., \& Yi, S., K. 2004, \apjs, in
press (astro-ph/0409024)

\bibitem[De Medeiros et al.(1996)]{DeMedeiros96}
De Medeiros, J.R., Da Rocha, C., and Mayor, M.~1996, \aap~314, 499.

\bibitem[De Medeiros et al.(1997)]{DeMedeiros97}
De Medeiros, J.R., Do Nascimento, J., and Mayor, M.~1997, \aap~317, 701.

\bibitem[De Medeiros \& Udry(1999)]{DeMedeiros99}
De Medeiros, J.R., and Udry, S.~1999 (DU99), \aap~346, 532.

\bibitem[Duncan(1981)]{Duncan81}
Duncan, D.K.~1981, \apj~248, 651.

\bibitem[Duquennoy \& Mayor(1991)]{Duquennoy1991}
Duquennoy, A., and Mayor, M.~1991, \aap~248, 485.

\bibitem[ESA(1997)]{HIP97}
ESA 1997, The Hipparcos and Tycho Catalogues, ESA SP-1200.

\bibitem[Fekel et al(2001)]{Fekel2001}
Fekel, F. et al~2001, \aj 122, 991.

\bibitem[Girardi et al.(2000)]{Girardi2000}
Girardi, L., Bressan, A., Bertelli, G. \& Chiosi, C~2000, \aaps~141, 371.

\bibitem[Hummel et al.(1993)]{Hummel1993}
Hummel, C.A.~et al.~1993, \aj~106, 2486.

\bibitem[Hummel et al.(1995)]{Hummel1995}
Hummel, C.~et al.~1995, \aj~110, 376.

\bibitem[Hummel et al.(1998)]{Hummel1998}
Hummel, C.~et al.~1998, \aj~116, 2536.

\bibitem[Hummel et al.(2001)]{Hummel2001}
Hummel, C.~et al.~2001, \aj~121, 1623.

\bibitem[Hummel et al.(2003)]{Hummel2003}
Hummel, C.~et al.~2003, \aj~125, 2630.

\bibitem[Hut(1981)]{Hut81}
Hut, P.~1981, \aap~99, 126.

\bibitem[Iben(1991)]{Iben91}
Iben, I.~1991, \apjs~76, 55.

\bibitem[Latham(1992)]{Latham1992}
 Latham, D.\ W.\ 1992, in IAU Coll.\ 135, Complementary Approaches to
Double and Multiple Star Research, ASP Conf.\ Ser.\ 32, eds.\ H.\ A.\
McAlister \& W.\ I.\ Hartkopf (San Francisco: ASP), 110

\bibitem[L\`ebre et al.(1999)]{Lebre99}
Lebre, A., De Laverny, P., De Medeiros, J., Charbonnel, C., and Da Silva, L.~1999,
\aap~345, 936.

\bibitem[Lejeune, Cuisinier, \& Buser(1998)]{Lejeune1998}
 Lejeune, T., Cuisinier, F., \& Buser, R. 1998, \aaps, 130, 65

\bibitem[Mallik et al.(2003)]{Mallik2003}
Mallik, S., Parthasarathy, M, and Pati, A.~2003, \aap~409, 251.

\bibitem[Mermilliod \& Mermilliod(1994)]{Mermilliod1994}
 Mermilliod, J.-C., \& Mermilliod M. 1994, Catalogue of Mean UBV Data
on Stars, (New York: Springer)

\bibitem[Mozurkewich et al.(1991)]{Mozurkewich91}
Mozurkewich, D.~et al.~1991, \aj~101, 2207.

\bibitem[Nordstr\"om et al.(2004)]{Nordstrom2004}
 Nordstr\"om, B., Mayor, M., Andersen, J., Holmberg, J., Pont, F.,
J\"orgensen, B.\ R., Olsen, E.\ H., Udry, S., \& Mowlavi, N. 2004,
\aap, 418, 989

\bibitem[Parsons(2004)]{Parsons2004}
 Parsons, S.~2004, \aj 127, 2915.

\bibitem[Pickles(1998)]{Pickles98}
Pickles, A.~1998, \pasp~110, 863.

\bibitem[Press et al.(1992)]{Press92}
Press, W.H., Teukolsky, S.A., Vetterling, W.T., and Flannery, B.P.~1992,
Numerical Recipes in C: The Art of Scientific Computing, Second Edition,
Cambridge University Press.

\bibitem[Queloz et al.(1998)]{Queloz98}
Queloz D., Allain S., Mermillod J.-C., Bouvier J., Mayor M.~1998,
\aap~335, 183.

\bibitem[Reiners \& Schmitt(2003)]{Reiners2003}
 Reiners, A., \& Schmitt, J.\ H.\ M.\ M. 2003, \aap, 398, 647

\bibitem[Shorlin et al.(2002)]{Shorlin2002}
 Shorlin, S.\ L.\ S., Wade, W.\ A., Donati, J.-F., Landstreet, J.\ D.,
Petit, P., Sigut, T.\ A.\ A., \& Strasser, S. 2002, \aap, 392, 637

%%\bibitem[Tokovinin(1997)]{Tokovinin97}
%%Tokovinin, A.~1997, \aaps~124, 75.

\bibitem[Torres et al.(2002)]{Torres2002}
Torres, G., Boden, A., Latham, D., Pan, M., and Stefanik, R.~2002, \aj~124, 1716.

\bibitem[Verbunt \& Phinney(1995)]{Verbunt95}
Verbunt, F.~and Phinney, E.~1995, \aap~296, 709.

%%\bibitem[Worley \& Douglass(1997)]{WDS97}
%%Worley, C.~and Douglass, G.~1997, \aaps~125, 523.

\bibitem[Yi et al.(2001)]{Yi2001}
 Yi, S., Demarque, P., Kim, Y.-C., Lee, Y.-W., Ree, C.\ H., Lejeune,
T., \& Barnes, S. 2001, \apj, 136, 417

\bibitem[Yi et al.(2003)]{Yi2003}
 Yi, S., Kim, Y.-C., \& Demarque, P. 2003, \apjs, 144, 259

\bibitem[Zahn(1977)]{Zahn77}
Zahn, J.-P.~1977, \aap~57, 383.

\bibitem[Zucker \& Mazeh(1994)]{ZM94}
Zucker, S. \& Mazeh, T.~1994, \apj~420, 806.


\end{thebibliography}
\end{document}